\documentclass[review]{elsarticle}

\usepackage{lineno}

\usepackage[lmargin=.75in,rmargin=.75in,tmargin=1.in,bmargin=1in]{geometry}

\usepackage[hyphens]{url}
\usepackage[colorlinks,citecolor=blue,urlcolor=blue]{hyperref}

\usepackage{url}

\usepackage{breakurl}
\usepackage{float}
\usepackage{caption}

\usepackage{siunitx}
\usepackage{xcolor}

\usepackage{bm}
\usepackage{amsfonts,amssymb,amsmath}
\usepackage{algorithm}
\usepackage[noend]{algpseudocode}

\newtheorem{thm}{Theorem}

\newdefinition{rmk}{Remark}
\newproof{pf}{Proof}
\newproof{pot}{Proof of Theorem \ref{thm2}}

\journal{arXiv}

\biboptions{authoryear}

\begin{document}
\nolinenumbers

\begin{frontmatter}

\title{ Quantifying Uncertainty with a Derivative Tracking SDE Model and Application to Wind Power Forecast Data} 

\author[1]{Renzo Caballero}
\ead{Renzo.CaballeroRosas@kaust.edu.sa}
\author[2]{Ahmed Kebaier}
\ead{kebaier@math.univ-paris13.fr}
\author[3]{Marco Scavino\corref{cor1}}
\ead{mscavino@iesta.edu.uy}
\author[1,4]{Ra\'ul Tempone}
\ead{raul.tempone@kaust.edu.sa, tempone@uq.rwth-aachen.de}

\cortext[cor1]{Corresponding author}
\address[1]{Computer, Electrical and Mathematical Sciences and Engineering Division (CEMSE), King Abdullah University of Science and Technology (KAUST), Thuwal 23955-6900, Saudi Arabia}
\address[2]{Universit\'e Sorbonne Paris Nord, LAGA, CNRS, UMR 7539, F-93430, Villetaneuse, France}
\address[3]{Instituto de Estadistica (IESTA), Universidad de la Rep\'ublica, 11200 Montevideo, Uruguay}
\address[4]{Alexander von Humboldt Professor in Mathematics for Uncertainty Quantification, RWTH Aachen University, 52702 Aachen, Germany}

\begin{abstract}

We develop a data-driven methodology based on parametric It\^{o}'s Sto\-chas\-tic Differential Equations (SDEs) to capture the real asymmetric dynamics of forecast errors. 
Our SDE framework features time-derivative tracking of the forecast, time-varying mean-reversion parameter, and an improved state-dependent diffusion term. 
Proofs of the existence, strong uniqueness, and boundedness of the SDE solutions are shown under a principled condition for the time-varying mean-reversion parameter. Inference based on approximate likelihood, constructed through the moment-matching technique both in the original forecast error space and in the Lamperti space, is performed through numerical optimization procedures. We propose another contribution based on the fixed-point likelihood optimization approach in the Lamperti space. 

All the procedures are agnostic of the forecasting technology, and they enable comparisons between different forecast providers. We apply our SDE framework to model historical Uruguayan normalized wind power production and forecast data between April and December 2019. Sharp empirical confidence bands of future wind power production are obtained for the best selected model. 
\end{abstract}


\end{frontmatter}


\section{Introduction} \label{Section_1}
 
In this work, we develop a methodology for analyzing a kind of data, often available in real-world problems, that consists of historical observations and their forecasts. Data-driven parametric stochastic differential equations (SDEs), whose solution defines a stochastic process, are the tool chosen to model the forecast errors. \\ This resultant stochastic process describes the time evolution dynamics of forecast errors while capturing properties such as a correlation structure and the inherent asymmetry. The model we propose is agnostic of the forecasting technology and serves to complement forecasting procedures by providing a data-driven stochastic forecast. Hence, we can evaluate forecasts according to their performance, and we can compare different forecasting technologies. \\ Most notably, we set up the ability to sample in the path space of the observed phenomena given a deterministic forecast $ \{p_t, \, t \in [0, T] \}$. Future simulated paths using Monte Carlo methods, as well as the analytic form of the proposed SDE, can be used in optimal control problems. \\ 
As a motivating application to show the proposed SDE framework at work, we consider probabilistic wind power forecasting.

Some interesting works have been devoted to probabilistic forecasting related to renewable energies based on stochastic differential equations, among them (\cite{mozuma}) and (\cite{elkate}) on wind power forecast, (\cite{immm}) and (\cite{bggk}) on forecasts of solar irradiance.
Here, we propose an improved model featuring time derivative tracking of the forecast, time-dependent mean reversion, modified diffusion, and non-Gaussian approximations. We apply the model to Uruguayan wind power forecasts together with historical wind power production data pertaining to the year 2019. \\

The rest of the paper is organized as follows. In Section \ref{Section_2}, we introduce the significant steps for constructing data-driven models for the normalized forecast process based on stochastic differential equations resulting in time-inhomogeneous generalizations of Jacobi type processes with mean-reversion. More precisely, we develop a modified drift for the model that incorporates the derivative of the available forecast so that the corresponding model is centered around $ \{p_t, \, t \in [0, T] \}$. Up to our knowledge, this is a new contribution that allows to kill efficiently the bias in the statistical inference problem. The application of the Lamperti transform with unknown parameters in Section \ref{Section_3} leads to model the forecast error through a stochastic differential equation with a unit diffusion coefficient. In Section \ref{Section_4}, we write down the expressions for the likelihood functions of the forecast error in its original space and the Lamperti space. We also derive tractable approximations of the likelihood functions based on the moment-matching technique. Section \ref{Section_4} concludes with the description of the optimization algorithms to compute approximate maximum likelihood estimates in the original forecast error space and in the Lamperti space.  In the latter case, the optimization step involves the use of a fixed-point approach that makes this procedure more stable than working with the raw data (See Algorithm \ref{alg1}). Up to our knowledge, it is a new approach to optimize efficiently the likelihood in a numerical stable and robust way. We also expand the model comprising an initial transition from the time the forecast is performed to the time of the first forecast. This generalization is relevant for applications since it allows the user to quantify forecast uncertainty from the beginning of every future period in an optimal way.
In Section \ref{Section_5}, we first describe the main characteristics of a real data set encompassing the normalized wind power production in Uruguay between April and December 2019, with the most accurate predictions, as highlighted in our posterior analysis, performed by one out of the three sources of forecast providers. Then, we apply our proposed numerical estimation procedures to the Uruguay wind and forecast dataset, comparing two alternative models with and without the derivative tracking drift component to assess the performance of the three different forecast providers. Our numerical results confirm that the latter is the best candidate model. Section \ref{Section_6} concludes the paper. 
The proofs of the existence, strong uniqueness, and boundedness of the SDE solutions used to model normalized wind power production and its forecast error are given in the Appendix.



\section{Data-driven stochastic differential equation models} \label{Section_2}

We build a type of phenomenological model for the normalized forecasts of an observable phenomena that, in its most general form, is a stochastic process $X = \{X_t, \, t \in [0, T] \}$  defined by the following stochastic differential equation (SDE):
\begin{equation}
  \left\{
  \begin{array}{@{}rl@{}}
    dX_t \!\!\!&=  a(X_t; p_t, \dot{p}_t,\bm{\theta}) \,dt + b (X_t; p_t, \dot{p}_t, \bm{\theta} ) \,dW_t\,, \;\; t \in [0,T]  \\
     X_0  \!\!\!&=  x_0 \in [0,1],
  \end{array}\right. \label{eq:main}
\end{equation}
where
\begin{itemize}
\item $a(\cdot; p_t, \dot{p}_t, \bm{\theta}): [0,1] \to \mathbb{R} $  denotes a drift function,
\item $b(\cdot; p_t, \dot{p}_t, \bm{\theta}): [0,1] \to \mathbb{R}^+ $  a  diffusion function,
\item $\bm{\theta}$ is a vector of unknown parameters,
\item $(p_t)_{t \in [0,T]}$ is a time-dependent deterministic function \\ $[0,1]$-valued and $ (\dot{p}_t)_{t \in [0,T]}$ is its time derivative,
\item $\{W_t, \, t \in [0,T] \}$ is a standard real-valued Wiener process.
\end{itemize}

In this work, $(p_t)_{t \in [0,T]}$ is to be considered a deterministic forecast for the normalized data, which is provided by an official source. 

Our goal is to achieve a specification of the model (\ref{eq:main}) to follow the available normalized forecasts closely while ensuring its unbiasedness with respect to the forecast.

\subsection{Data constraints} \label{Data_Constraints}

Let $(p_t)_{t \in [0,T]}$ be the available prediction function for the normalized observed real data, which is the main input to this approach. Most of previous studies dealing with the problem of error forecast quantification through S.D.E. models proposed a drift of the form $-\theta_t (X_t - p_t)$ (see e.g. \cite{elkate} and \cite{bggk} with $\theta_t\equiv a$ a positive constant). However, it is clear that such a choice leads  the model to revert to 
$$
\mathbb E[X_t]= e^{-\int_0^t\theta_sds}\Big(\mathbb E[X_0]+ \int_0^te^{\int_0^s\theta_udu}p_s\theta_sds\Big),
$$
which is not the natural value that one would expect for a forecast  probabilistic model (See Remark \ref{rmk:mean_revert} below). To overcome this crucial data constraint, we introduce a time-dependent drift function that features the expected mean-reverting property as well as derivative tracking:
\begin{equation}
a(X_t; p_t, \dot{p}_t, \bm{\theta}) = \dot{p}_t  - \theta_t (X_t - p_t),  \label{drift:meanrev-derivtrack}
\end{equation} 
where $ (\theta_t)_{t \in [0,T]} $ is a positive deterministic function, whose range depends on $\bm{\theta}$, as will be explained shortly.
More precisely, the normalized  forecast process $X_t$, modeled as solution to the It\^{o} stochastic differential equation (\ref{eq:main}) with the drift specified in (\ref{drift:meanrev-derivtrack}) satisfies now $\mathbb{E} \left[X_t\right] = p_t$, given that $  \mathbb{E}\left[X_0\right] = p_0$, since by It\^{o}'s lemma we get 
\begin{align*}
e^{\int_{0}^{t} \theta_s ds}  X_t - X_0 & = \int_{0}^{t}  (\dot{p}_s + \theta_s p_s ) e^{\int_{0}^{s} \theta_u du} ds 
\nonumber \\  
& + \int_{0}^{t}  b (X_s; p_s, \dot{p}_s, \bm{\theta} )  e^{\int_{0}^{s} \theta_u du} d W_s.  \label{eq:tomeanX}
\end{align*}
and consequently
\begin{equation}
\mathbb{E} \left[X_t\right] =  e^{ - \int_{0}^{t} \theta_s ds} \left(  \mathbb{E}\left[ X_0\right] + p_t \, e^{ \int_{0}^{t} \theta_s ds} - p_0 \right) = p_t.
\label{eq:meanX}
\end{equation}

 At this stage, this novel process model satisfies the two main following properties: 
\begin{itemize}
\item it reverts to its mean $p_t$, with a time-varying speed $ \theta_t$ that is proportional to the deviation of the process $X_t$ from its mean,
\item it tracks the time derivative $\dot{p}_t$, 
\end{itemize} 
which to the best of our knowledge is an original contribution for the study of forecast modeling problem using stochastic differential equations.

\begin{figure}[H]
\centering
  \includegraphics[width=0.485\textwidth]{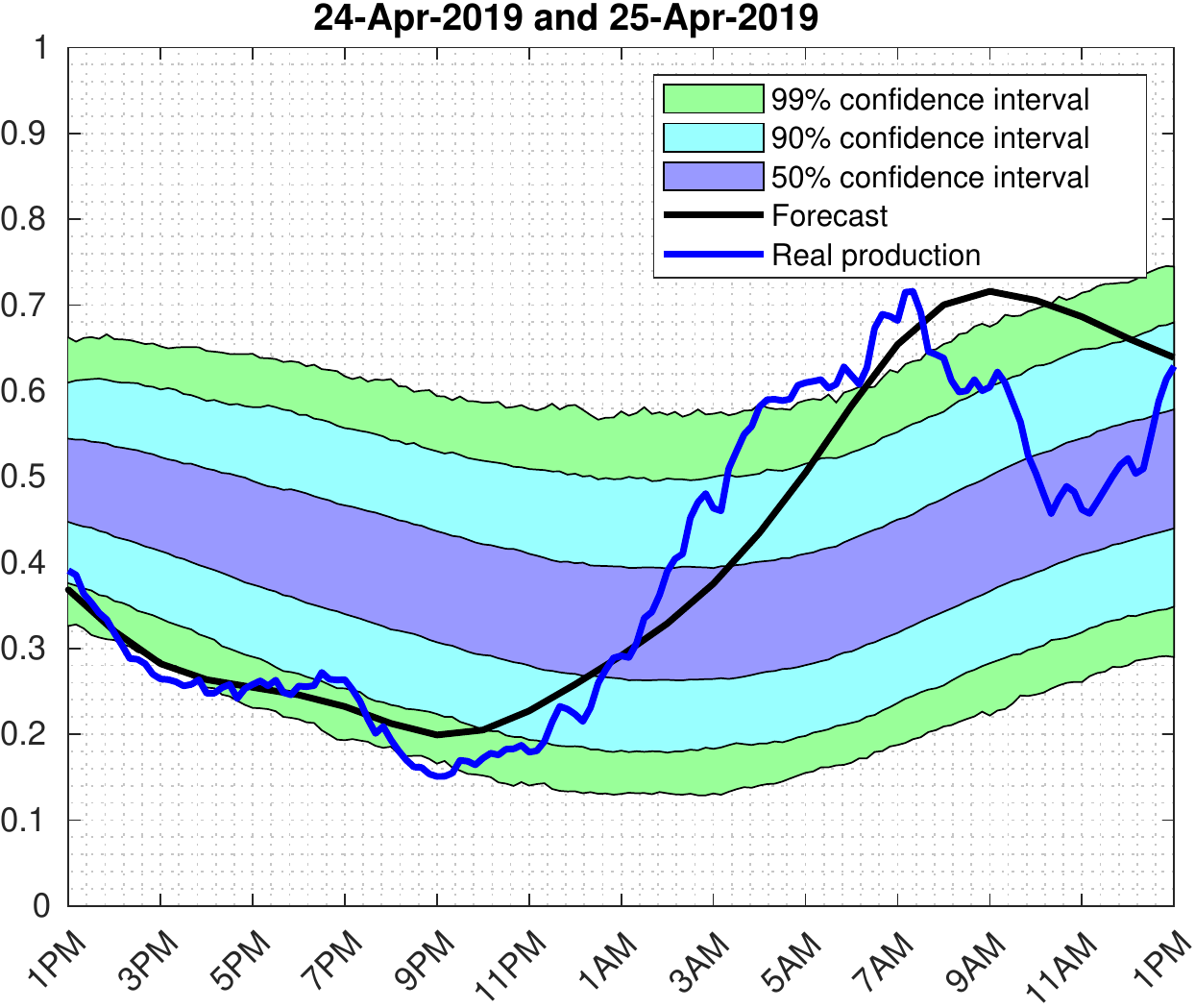}\quad\includegraphics[width=0.485\textwidth]{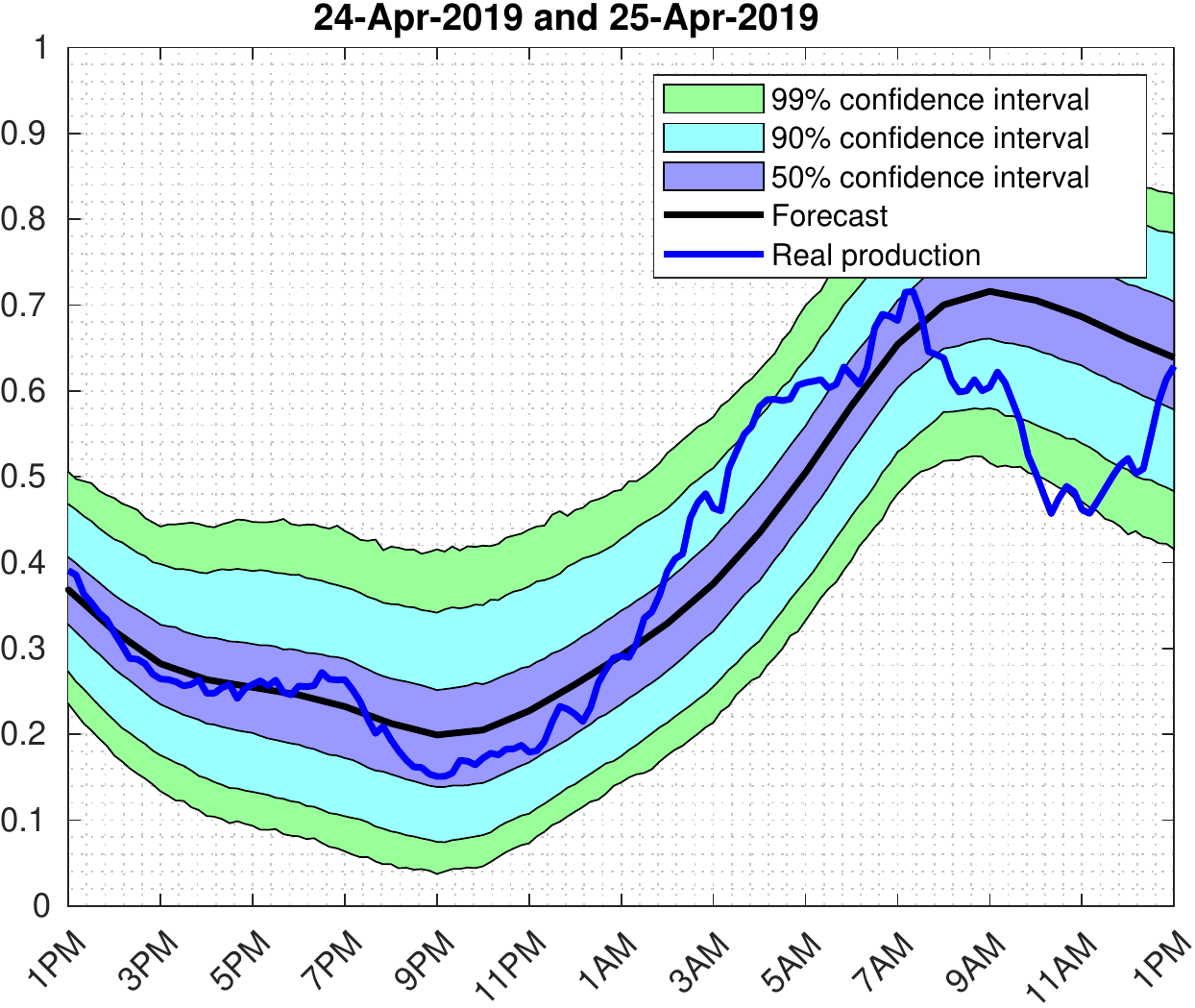}
  \caption{Pointwise confidence bands fitted, for the same daily segment, through diffusion models without derivative tracking (plot on the left) and with derivative tracking (plot on the right).}
  \label{fig:derivative_tracking}
\end{figure}

\begin{rmk}\label{rmk:mean_revert}
Observe that a mean-reverting model without derivative tracking shows a delayed path behavior. For instance, consider the diffusion model (\ref{eq:main}) with $a(X_t; p_t, \bm{\theta}) = - \theta_0 (X_t - p_t)\,, \theta_0 > 0$. In this case, given  $ \mathbb{E} \left[X_0\right] = p_0$, the diffusion has mean $\mathbb{E} \left[X_t\right] = p_t - e^{- \theta_0 t } \int_0^t \dot{p}_s  e^{\theta_0 s} ds$. Figure (\ref{fig:derivative_tracking}) illustrates, on the wind power forecast data, how different behave the estimated confidence bands for two diffusion models with and without derivative tracking, fitting the same daily segment.
\end{rmk}

The observable phenomena measurements and forecasts data are properly normalized. For example, the forecast and production wind power data of Uruguay are normalized with respect to the installed power capacity during the period of observation. Thus, the mean-reverting level lies in $[0,1]$, and the process $X_t$  must take values in the same interval, a requirement that is not automatically fulfilled through the derivative tracking. To impose that the state space of $X_t$ is $[0,1]$, we may choose a convenient diffusion term, and require that the time-varying parameter $ \theta_t$ satisfies an ad-hoc condition.
 
Let $\bm{\theta} = (\theta_0,\alpha)$, and choose a state-dependent diffusion term that avoids the process exiting from the range $[0,1]$ as follows:
  \begin{equation}
    b (X_t; \bm{\theta} )= \sqrt{2 \alpha \theta_0 X_t (1 - X_t)}
  \end{equation}
  where $\alpha >0$ is an unknown parameter that controls the path variability. This diffusion term belongs to the Pearson diffusion family and, in particular, it defines a Jacobi type diffusion. It is useful to recall that \cite[440]{foso} a Pearson diffusion is a stationary solution to a stochastic differential equation of the form \begin{equation}
    dX_t = - \theta (X_t - \mu) dt + \sqrt{2 \theta \left(a X_t^2 + b X_t + c\right)} d W_t
    \label{eq:poly_diff}
  \end{equation}
where $\theta>0$, and $a$, $b$, and $c$ are parameters such that the square root is well defined when $X_t$ is in the state space. These parameters, together with $\mu$, the mean of the invariant distribution, determine the state space of the diffusion as well as the shape of the invariant distribution.

An exhaustive classification of the (stationary) Pearson diffusions is presented in \cite[440-443]{foso} where, in particular, it is discussed the case $a < 0$ and $b(x; \bm{\theta}) = \sqrt{2 a \theta x (x-1)}$, where the invariant distribution is a Beta distribution with parameters $\left( \frac{\mu}{-a}, \frac{1 - \mu}{-a} \right),$ that leads to the well-known Jacobi diffusions, so-called because the eigenfunctions of the infinitesimal generator of these processes are the Jacobi polynomials (see, for example, \cite[2860-2861]{leph}). 

It is worth mentioning that Jacobi diffusions have been successfully applied in several disciplines, among them finance (see (\cite{vago}) and references therein) and neuroscience (\cite{dotala}).

However, a distinctive feature in our proposed model 
\begin{equation}
  \left\{
  \begin{array}{@{}rl@{}}
    dX_t \!\!\!&= (\dot{p}_t  - \theta_t (X_t - p_t) ) dt +\sqrt{2 \alpha \theta_0 X_t (1 - X_t)} dW_t\,, \;\; t \in [0,T]  \\
   X_0  \!\!\!&=  x_0 \in [0,1] \,,
 \end{array}\right.  \label{ourmodel}
\end{equation} 
is that the drift term contains the time-varying parameter $\theta_t$, rendering the solution $X_t$ of (\ref{ourmodel}) to a non-stationary and time-inhomogeneous process. To ensure that the process $X_t$ is the unique strong solution of (\ref{ourmodel}) for all $t \in [0,T]$ with state space $[0,1]$ a.s., the mean-reversion time-varying parameter must satisfy the condition:

\begin{equation}
\theta_t\geq \max\left(\frac{\alpha\theta_0+\dot p_t}{1-p_t},\frac{\alpha\theta_0-\dot p_t}{p_t}\right)\tag{B}. 
\end{equation} \label{condB}
The proof of this theoretical statement is presented in the Appendix. 

\begin{rmk}
Condition (B) shows that the time-varying parameter $\theta_t$ becomes unbounded when $p_t = 0$ or $p_t = 1$. Therefore, we consider the following truncated prediction function
\begin{equation}
p_t^{\epsilon}=\begin{cases}
\epsilon&\quad\text{if}\quad p_t<\epsilon\\
p_t&\quad\text{if}\quad\epsilon\leq p_t<1-\epsilon\\
1-\epsilon&\quad\text{if}\quad p_t\geq1-\epsilon
\end{cases}
\label{corrforecast}
\end{equation}
that satisfies $p_t^{\epsilon} \in [\epsilon, 1 - \epsilon]$ for any $0 < \epsilon < \frac{1}{2}$ and $t \in [0,T]$, providing that $\theta_t$ is bounded for every $t \in [0,T]$. \\
For any forecast dataset,  a small $\epsilon >0$ needs to be specified to define the truncated prediction function fulfilling the above condition.
\end{rmk}

From now on, we will keep the notation $p_t$ to denote the truncated prediction function (\ref{corrforecast}), unless specified otherwise. 

\subsection{A model specification for the forecast error}
 After applying to (\ref{ourmodel}) the simple change of variables $$V_t = X_t - p_t \,,$$ we may introduce the following model for the normalized forecast error:
\begin{equation}
  \left\{
  \begin{array}{@{}rl@{}}
    dV_t \!\!\!&=  - \theta_t V_t dt + \sqrt{2 \alpha \theta_0 (V_t +p_t ) (1-V_t-p_t)} dW_t , \;\; t \in [0,T]  \\
     V_0  \!\!\!&=  v_0 \in [- p_0,1 - p_0].
  \end{array}\right. \label{VtSDE}
\end{equation}



\section{State independent diffusion term: Lamperti transform} \label{Section_3}

Our model (\ref{VtSDE}) for the forecast error has a diffusion term that depends on the state variable $V_t$. Under the conditions that permit the use of It\^{o}'s formula on a well-chosen transformation of the process $V$, John Lamperti (\cite{lamp}) first showed that the transformed process is again a diffusion process that is solution to a SDE with unit coefficient for the diffusion term. 
The vast literature nowadays refers to this result as the so-called Lamperti transform (see, for example, \cite[40--41]{iacus1}; \cite{moma}; \cite[199--203]{pani}; \cite[98--100]{saso}), which is a basic tool to obtain a SDE for the transformed process whose diffusion term does not depend anymore on the state variable. A remarkable effect of removing the state dependency from the random noise term is to increase the numerical stability of the simulated paths of the transformed process. 
For this reason, some estimation methods of the unknown parameters of non-linear SDE models incorporated the Lamperti's change of variable as part of a more complex approximation procedure (for example, in the case of one-dimensional diffusions, the local linearization method in  \cite{shoz}, or the expansion method in \cite{ait}, later extended to time-inhomogeneous SDEs in \cite{eglix}).

We consider the following Lamperti transform with unknown parameters
\begin{align}
\begin{split}
Z_t = h(V_t, t; \bm{\theta} ) & = \frac{1}{\sqrt{2 \alpha \theta_0 }} \int \frac{1}{\sqrt{(v + p_t)(1 - v - p_t)}} dv \Bigg{\vert}_{v = V_t}  \\
& = - \sqrt{ \frac{2}{ \alpha \theta_0 }} \arcsin (\sqrt{ 1 - V_t - p_t}) \label{eq:LampZ}
\end{split}
\end{align}
that, after applying It\^{o}'s formula on $h(V_t, t; \bm{\theta} )$, leads to the following SDE with state independent unit diffusion term
\begin{equation}
dZ_t = \Bigg[   \frac{\dot{p}_t}{ \sqrt{2 \alpha \theta_0 (V_t + p_t)(1 - V_t - p_t)}}  
+ \frac{- \theta_t V_t}{ \sqrt{2  \alpha \theta_0 (V_t + p_t)(1 - V_t - p_t)}}  - \frac{1}{4} \frac{\sqrt{2 \alpha \theta_0} 
\left( 1 - 2 (V_t + p_t)\right)}{\sqrt{(V_t + p_t)(1 - V_t - p_t)}}  \Bigg] dt + d W_t . \label{eq:stindepSDE}
\end{equation}
After replacing $V_t = 1 - p_t - \sin^2 \left(- \sqrt{ \frac{ \alpha \theta_0}{2} } Z_t \right) $ in (\ref{eq:stindepSDE}), we obtain that the  process $Z_t$ satisfies the SDE
{\small
\begin{align}
\begin{split}
dZ_t  =& \left[ \frac{ \dot{p}_t  - \theta_t  \left(1 - p_t - \sin^2 \Big(- \sqrt{ \frac{ \alpha \theta_0}{2} } Z_t \Big) \right) }{\sqrt{2 \alpha \theta_0} \cos\Big(- \sqrt{ \frac{ \alpha \theta_0}{2} } Z_t \Big)  \sin\Big(- \sqrt{ \frac{ \alpha \theta_0}{2} } Z_t \Big)} 
- \frac{1}{4}  \frac{\sqrt{2 \alpha \theta_0} \left(1 - 2 \cos^2 \Big(- \sqrt{ \frac{ \alpha \theta_0}{2} } Z_t \Big) \right) }{\cos\Big(- \sqrt{ \frac{ \alpha \theta_0}{2} } Z_t \Big)  \sin\Big(- \sqrt{ \frac{ \alpha \theta_0}{2} } Z_t \Big)} \right] dt + d W_t \\
 = &\left[  \frac{  2  \dot{p}_t - \theta_t (1 - 2 p_t)  + (\alpha \theta_0 - \theta_t) \cos(- \sqrt{2 \alpha \theta_0 } Z_t) }{\sqrt{2 \alpha \theta_0} \sin{(- \sqrt{2 \alpha \theta_0} Z_t)}}  \right] dt + d W_t.  \label{eq:stindepSDE2}
\end{split}
\end{align}}

A visual summary of the effect of the Lamperti transform can be appreciated later in Section \ref{Section_5}, Figure  (\ref{fig:LP_transitions}), where we can see in the wind power application how the forecast error transition histograms (without curtailment) modify in comparison with Figure (\ref{fig:error_transitions}).

The shape of the forecast error transition histograms after applying the Lamperti transform has similarities with the Gaussian distribution, motivating toward the use of Gaussian-like approximations of the unknown density transition functions of the process $Z_t$.

\begin{rmk}
In general, when we introduce a diffusion term of a Jacobi type process, on one side the Beta density function appears as a natural candidate to deal with the asymmetric trait of the data. On the other side, the advantage of the Lamperti transform is contributing to remove asymmetry in data, allowing the use of the Gaussian density as surrogate for the unknown transition density function. Moreover, this obtained Gaussian distribution supports the validity of the choice of our model diffusion coefficient given by (\ref{eq:poly_diff}). 
\end{rmk}



\section{ Likelihood functions of the forecast error data and optimization algorithm} \label{Section_4} 

\subsection{Likelihood in the $V-$space}

Suppose that any of $M$ non-overlapping paths of the con\-tin\-u\-ous-time It\^{o} process $V = \{ V_t, t  \in [0,T] \}$, each one starting at a different time $t_j$ with $j = 1, \dots, M$, is sampled at $N + 1$ equispaced discrete points with given length interval $\Delta$. Let $ V^{M,N + 1}=\left\{ V_{t_1^{N + 1}} , V_{t_2^{N + 1}} ,\ldots , V_{t_M^{N + 1}} \right\}$ denote this random sample, with $V_{t_j^{N + 1}} =\left\{ V_{t_j + i \Delta}\,, i = 0, \ldots, N \right\}$.

Let $\rho(v \vert v_{j, i-1} ; \bm{\theta})$ be the conditional probability density of $V_{t_j + i \Delta} \equiv V_{j, i}$ given $V_{j, i-1} = v_{j, i-1}$ evaluated at $v$, where $\bm{\theta} = (\theta_0, \alpha)$ are the unknown model parameters.

The It\^{o} process $V$ defined by the SDE (\ref{VtSDE}) is Markovian, and the likelihood function of the sample $ V^{M,N + 1}$ can be written as the following product of transition densities:  
\vspace{-0.1cm}
\begin{equation}
\mathcal{L}\left(\bm{\theta}; V^{M,N +1}\right) = \prod\limits_{j=1}^M \left\{ \prod\limits_{i=1}^N \rho \left( {V_{j, i}| V_{j, i-1}} ; p_{[t_{j,  i-1}, t_{j , i} ]},  \bm{\theta} \right)    \right\},
\label{likelihood}
\end{equation}
where $t_{j ,i} \equiv  t_j + i \Delta$ for any $j = 1, \ldots, M$ and $i = 0, \ldots, N$. 

\begin{rmk} \label{remdelta}
In the last subsection of this section, we will extend the statistical model (\ref{likelihood}) by adding the transition that occurs during the time interval, say of length $\delta$, between the epoch when the forecast is done and the first epoch (1 pm) of each day-ahead forecast. 
To this purpose, the likelihood function (\ref{likelihood}) must include for any of the $M$ paths an additional factor, say $\rho_0 (V_{j, 0}|V_{j, -\delta};\bm{\theta},\delta)$, expressing the conditional density of the early transition. The parameter $\delta$ can be calibrated together or after the estimation of $\bm{\theta}$, suggesting an optimal time for the scheduling of the forecasts.
\end{rmk}
 
The exact computation of the likelihood (\ref{likelihood}) relies on the availability of a closed-form expression for the transition densities of $V$ that, on the basis of the Markovian property of $V$, are characterized for $ t_{j, i-1} < t < t_{j,i}$,  as solutions of the Fokker-Planck-Kolmogorov equation (\cite[36]{iacus1}; \cite[61-68]{saso}):
\begin{align}
\frac{ \partial f }{\partial t } & \rho(v ,t \vert v_{j,i-1} ,  t_{j,i-1} ; \bm{\theta} )= - \frac{\partial}{ \partial v} (- \theta_t v \, \rho(v ,t \vert v_{j,i-1} ,  t_{j,i-1} ; \bm{\theta} ) ) \nonumber \\
& + \frac{1}{2} \frac{\partial^2}{ \partial v^2} ( 2 \theta_0 \alpha (v+ p_t) (1 - v- p_t) \, \rho(v ,t \vert v_{j,i-1} ,  t_{j,i-1} ; \bm{\theta} ) ),   \nonumber \\  &  v \in (-1 ,1) , \:\:t > 0 \label{eq:fpk}
\end{align}
subject to the initial conditions $\rho(v , t_{j, i-1} ; \bm{\theta} ) = \delta(v - V_{j, i-1}) \,,$ where $ \delta(v - V_{j, i-1})$ is the Dirac-delta generalized function centered at $ V_{j, i-1}\,.$

However, closed-form solutions to initial-boundary value problems for time-inhomogeneous diffusions can be obtained only in a few cases (see, for example, \cite[Section 3.1]{eglix}). In our case, solving numerically (\ref{eq:fpk}) for the transition densities of the process $V$ at every transition step is computationally expensive. 
Several numerical techniques have been devised to obtain estimates for the unknown parameters of continuous-time SDE models with discrete observations (see, for example, \cite{prewo} for likelihood-based inference techniques, \cite{Sor} for an estimating function approach). As explained in the next subsection, we have considered approximate likelihood methods, similar in spirit to \cite[Section 11.4]{saso}.

\subsection{Approximate likelihood  in the $V-$space} \label{moments_ODEs}

Gaussian approximations to the transition densities of nonlinear time-in\-ho\-mo\-ge\-ne\-ous SDEs are available through different algorithms \cite[Chapter 9]{saso}. However, as Figure \ref{fig:error_transitions} may suggest at first glance, the choice of a Gaussian density could be inadequate when straightly applied to approximate the transition density of the forecast error $V$ of the normalized wind power production. 

Therefore, we propose to use a surrogate transition density for $V$ other than Gaussian. The moments of the SDE model (\ref{VtSDE}) are then matched to the surrogate density moments. 

From (\ref{eq:meanX}), we have $m_1(t) \equiv \mathbb{E} \left[V_t\right] = e^{- \int_{t_{j,i-1}}^t \theta_s ds} \,\mathbb{E} \left[V_{t_{j,i-1}}\right]$, for any $t\in [t_{j,i-1}, t_{j, i}[$, $j = 1, \ldots, M$ and $i = 1, \ldots, N\,.$ 

For $m \geq 2$, using It\^o's lemma we derive
\begin{align}
\frac{d \mathbb{E}\left[ V^m_t\right]}{d t} & = - m ( \theta_t + (m-1) \alpha \theta_0 ) \mathbb{E}\left[ V^m_t\right]   \nonumber \\ 
& + m (m-1) \alpha \theta_0   (1 - 2 p_t) \mathbb{E}\left[ V_t^{m-1} \right] \nonumber \\ 
& + m (m-1) \alpha \theta_0 p_t (1 -p_t) \mathbb{E}\left[ V_t^{m-2} \right]  .
\end{align}

For any $t\in [t_{j,i-1}, t_{j, i}[$, the first two moments of $V$, $m_1(t)$ and $m_2(t) \equiv \mathbb{E}\left[V_t^2\right]$, can be computed by solving the following system
\begin{equation}
\begin{cases}
\frac{d  m_1 (t)}{d t} &=  - m_1(t)\theta_t   \\
\frac{d  m_2 (t)}{d t} &=  -2 (\theta_t +\alpha \theta_0) m_2(t) + 2 \alpha\theta_0 (1-2p_t)  m_1(t) + 2 \alpha\theta_0 p_t (1-p_t) 
\end{cases}
\label{Vtmom}
\end{equation}
with initial conditions $m_1(t_{j,i-1})= v_{j, i-1}$ and $m_2(t_{j,i-1})= v_{j, i-1}^2 \,.$

\subsubsection{Moment Matching}

A suitable candidate for a surrogate transition density of $V$ is a Beta distribution on a compact interval parameterized by two positive shape parameters, $\xi_1, \xi_2$. {Recall that the choice of the Beta proxy distribution is a natural choice as it is the invariant distribution of the Jacobi type processes.} 

For any $t\in [t_{j,i-1}, t_{j, i}[$, we approximate the transition densities of the process $V$ using a Beta distribution. We equal the first two central moments of $V$ with the corresponding moments of the Beta surrogate distribution on $[-1 + \epsilon,1 - \epsilon]$ with shape parameters $\xi_1, \xi_2$.

The shape parameters are given by
\begin{equation}
 \begin{split}
\xi_1(t) & = - \frac{(\mu_t + 1 - \epsilon)(\mu_t^2 + \sigma_t^2 - (1- \epsilon)^2)}{2 (1 - \epsilon) \sigma_t^2}, \\ 
\xi_2(t) & =  \frac{(\mu_t-1 + \epsilon )(\mu_t^2 + \sigma_t^2 - (1- \epsilon)^2)}{2 (1 - \epsilon) \sigma_t^2} , \label{param_transformed_beta}
 \end{split}
 \end{equation}
where $\mu_t = m_1 (t)$ and $\sigma_t^2= m_2 (t)- m_1 (t)^2\,.$ \\

The approximate log-likelihood $\tilde{\ell}(\cdot ; v^{M, N+1})$ of the observed sample $v^{M, N+1}$ can be expressed as
\begin{equation}
 \tilde{\ell} \left(\bm{\theta}; v^{M,N +1}\right) 
 = \sum_{j=1}^M \sum_{i=1}^N \log  \Bigg\{ \frac{1}{2(1 - \epsilon)} \frac{1}{B(\xi_1(t_{j,i}^-), \xi_2(t_{j,i}^-))} \left( \frac{v_{j,i} + 1 - \epsilon}{2(1 - \epsilon)} \right)^{\xi_1(t_{j,i}^-) -1}  \left( \frac{1 - \epsilon - v_{j,i}}{2(1 - \epsilon)} \right)^{\xi_2(t_{j,i}^-) -1} \Bigg\},
\label{eq:loglikelihoodV}
\end{equation}
where the shape parameters $\xi_1(t_{j,i}^-)$ and $\xi_2(t_{j,i}^-)$, according to (\ref{param_transformed_beta}), depend on the limit quantities $\mu(t_{j,i}^-;\bm{\theta} )$ and $\sigma^2(t_{j,i}^-;\bm{\theta} )$ as $t\uparrow t_{j,i}$ that are computed solving numerically the initial-value problem (\ref{Vtmom}). $B(\xi_1,\xi_2)$ denotes the beta function.

\subsection{Approximate likelihood  in the $Z-$space} \label{moments_ODEs_Z}

The transition density of the process $Z$, which has been defined through the Lamperti transformation (\ref{eq:LampZ}) of $V$, can be conveniently approximated by a Gaussian surrogate density. 

The drift coefficient $a(Z_t; p_t, \dot{p}_t, \bm{\theta}) $ of the process $Z$ that satisfies (\ref{eq:stindepSDE2}) is nonlinear. After linearizing the drift around the mean of $Z$, $\mu_Z(t) \equiv \mathbb{E}\left[Z_t\right]$,  we obtain the following system of ODEs to compute, for any $t\in [t_{j,i-1}, t_{j, i}[$, the approximations of the first two central moments of $Z$, say  $\tilde{\mu}_Z(t) \approx \mathbb{E}\left[Z_t\right]$ and $\tilde{v}_Z(t) \approx \text{Var} \left[Z_t\right]$:
\begin{equation}
\begin{cases}
\frac{d \tilde{\mu}_Z (t)}{d t}&=  a\big( \tilde{\mu}_Z (t) ; p_t, \dot{p}_t, \bm{\theta} \big)   \\
\frac{d \tilde{v}_Z (t)}{d t}&= 2  a^{\prime} \big( \tilde{\mu}_Z (t) ; p_t, \dot{p}_t, \bm{\theta} \big) \tilde{v}_Z (t) + 1
\end{cases}
\label{eq:Ztmom}
\end{equation}
with initial conditions $\tilde{\mu}_Z(t_{j,i-1})= z_{j, i-1}$ and $\tilde{v}_Z(t_{j,i-1})= 0 \,,$ and where 
\begin{equation*}
a^{\prime} \left( \tilde{\mu}_Z (t) ; p_t, \dot{p}_t, \bm{\theta} \right)  
=   \frac{  (\alpha \theta_0 - \theta_t)  - \cos(\sqrt{2 \alpha \theta_0 } Z_t) [ \theta_t (1 - 2 p_t) - 2  \dot{p}_t ] }{\sin^2{(\sqrt{2 \alpha \theta_0} Z_t)}} \,.  
\end{equation*}
The approximate Lamperti log-likelihood $\tilde{\ell}_Z\left(\cdot ; z^{M, N+1}\right)$ for the observed sample $z^{M, N+1}$ is given by
\begin{equation}
\tilde{\ell}_Z \left(\bm{\theta}; z^{M,N +1}\right) 
= \sum_{j=1}^M \sum_{i=1}^N \log \left\{ \frac{1}{\sqrt{2 \pi \tilde{v}_Z(t_{j,i}^-; \bm{\theta})}} \exp \Bigg( -\frac{(z_{j,i} - \tilde{\mu}_Z(t_{j,i}^-;\bm{\theta} ))^2}{2 \tilde{v}_Z(t_{j,i}^-; \bm{\theta})} \Bigg) \right\},
\label{loglikelihoodZ}
\end{equation}
where the limits $\tilde{\mu}_Z(t_{j,i}^-;\bm{\theta} )$ and $\tilde{v}_Z(t_{j,i}^-;\bm{\theta} )$ are computed solving numerically the initial-value problem (\ref{eq:Ztmom}). 

\subsection{Algorithm for the approximate maximum likelihood estimations} \label{opt_sec}

{In this subsection, we aim to infer the model's parameters using optimization techniques. We start by finding an initial guess close enough to the optimal value, and from that point, start the optimization.}

\subsubsection{Initial guess}

To guarantee the good behave for our optimization algorithm, we aim to start the optimization as close as we can from the optimal parameters. We use {least square minimization} and {quadratic variation} over the data to find an initial guess $(\theta_0^*,\alpha^*)$.

\begin{itemize}

\item {Least square minimization:} We consider the observed data $v^{M,N+1}$ with length between observations $\Delta$, where $i\in\{0,\dots,N\}$ and $j\in\{1,\dots,M\}$. {For any $t\in [t_{j,i-1}, t_{j, i}[$, the random variable $(V_{j,i}|v_{j,i-1})$ has a conditional mean that can be approximated by the solution of the system}
\begin{equation*}
\begin{cases}
d  \mathbb{E} \left[ V \right](t)&=-\theta_t  \mathbb{E} \left[V \right](t) dt\\
 \mathbb{E} \left[V\right](t_{j,i-1})&=v_{j,i-1},
\end{cases}
\end{equation*}
in the limit $t\uparrow t_{j,i}$, i.e., $ \mathbb{E} \left[V\right](t_{j,i}^-)$. Then, the random variable $(V_{j,i} -  \mathbb{E} \left[V\right](t_{j,i}^-))$ has zero mean. If we assume that $\theta_t=c\in\mathbb{R}^+$ for all $t\in[t_{j,i-1},t_{j,i}[$, then $ \mathbb{E} \left[V\right](t_{j,i}^-)=v_{j,i-1}e^{-c\Delta}$. If we have a total of $M\times N$ transitions, we can write the regression problem for the conditional mean with $L^2$ loss function as
\begin{equation}
\begin{split}
\hat{c}&=\arg\min_{c\ \geq\ 0}\left[\sum_{j=1}^M\sum_{i=1}^{N}\left(v_{j,i}-  \mathbb{E} \left[V\right](t_{j,i}^-)\right)^2\right]\\
&\approx\arg\min_{c\ \geq\ 0}\left[\sum_{j=1}^M\sum_{i=1}^{N}\left(v_{j,i}-v_{j,i-1}(1-c\Delta)\right)^2\right].
\end{split}
\label{Eq-1}
\end{equation}
As Equation (\ref{Eq-1}) is convex in $c$, it is enough to verify the first order optimality conditions. 
It follows that
\begin{equation}
\hat{c}\approx\frac{\sum_{j=1}^M\sum_{i=1}^{N}v_{j,i-1}(v_{j,i-1}-v_{j,i})}{\Delta\cdot\sum_{j=1}^M\sum_{i=1}^{N}(v_{j,i-1})^2}.
\label{Eq-4}
\end{equation}
{We approximate $\theta_0$ by Equation (\ref{Eq-4}) setting $\theta_0^*=\hat{c}$.}

\item {Quadratic variation:} We approximate the quadratic variation of the  It\^{o}'s  process $V$,
$ \langle V\rangle _t=\int_0^t b(V_s; \bm{\theta}, p_s)^2 ds $, where  
\begin{equation*}
b(V_s; \bm{\theta}, p_s) = \sqrt{2 \alpha \theta_0 (V_s +p_s ) (1-V_s - p_s)},
\end{equation*}
with the discrete sum $\sum_{0< t_{j, i-1} \leq t}\left(V_{t_{j, i}} - V_{t_{j, i-1}}\right)^2$.\smallskip

As initial guess for the diffusion variability coefficient $\theta_0 \alpha$, we choose
\begin{equation}
\theta_0^*\alpha^*=\frac{\sum_{j=1}^M\sum_{i=1}^N(v_{j,i} - v_{j,i-1})^2}{2\Delta\cdot\sum_{j=1}^M\sum_{i=1}^N(v_{j,i}+p_{j,i})(1-v_{j,i}-p_{j,i})},
\label{Eq-2}
\end{equation}
where $\Delta$ is the length of the time interval between two consecutive measurements.
\end{itemize}

\subsubsection{Negative log-likelihood minimization in the $V-$space} \label{Sec:MinLH}

To find the optimal parameters, we minimize the negative log-likelihood (negative version of (\ref{eq:loglikelihoodV})) using the derivative-free function \textit{fminsearch} from MATLAB R2019b over the parameters $(\theta_0,\alpha)$. At each step of the iteration, we:
\begin{itemize}

\item Use the training dataset to find the SDE's first and second moments as explained in Subsection \ref{moments_ODEs}.
\item Match the proxy distribution moments with the SDE's moments.
\item Evaluate the negative log-likelihood using the training dataset.

\end{itemize}

\subsubsection{Negative log-likelihood minimization in the $Z-$space} \label{Sec:MinLHL}

Let $v^{M,N+1}$ be the observed data, and $h( v_{j,i},t_{j,i};\bm{\theta})$ the Lamperti transform of the observation $v_{j,i}$. As we can see in Section \ref{Section_4}, the transformed observations $z^{M,N+1}$ depend on the vector $\bm{\theta}$.\\
The problem of maximizing the approximated Lamperti log-likelihood (\ref{loglikelihoodZ}), i.e.,  $$\max_{\bm{\theta}}\tilde{\ell}_Z\left(\bm{\theta}; z^{M,N+1}\right),$$
is not totally defined as the data $z^{M,N+1}$ depend on $\bm{\theta}$. To address this issue, we propose to find a fixed point $\bm{\theta}^\star $ such that
\begin{equation}
\bm{\theta}^\star=\arg\max_{\bm{\theta}}\tilde{\ell}_Z\left(\bm{\theta};\{h( v_{j,i},t_{j,i};\bm{\theta}^\star )\}_{j=1,i=0}^{M,N}\right).
\label{FP}
\end{equation}
Thus, at a fixed point, the likelihood has a maximum for the transformed data set corresponding to that parameter value. The solution to \eqref{FP} is not available in closed form and therefore approximated numerically.

\begin{rmk} 
The optimization approach introduced in this subsection constitutes, up to our knowledge, a new proposal to get numerically robust and stable maximum likelihood estimates when applying the Lamperti transform to a diffusion process with unknown parameters.
\end{rmk}

\subsection{Model specification with the additional parameter $\delta$}

In real-world applications, the forecast error at time $t_{j,0}=0$ is not usually zero. According to the forecasts procedure, we may assume that there is a time in the past $t_{j,-\delta}<t_{j,0}$, such that the forecast error $V_{j,-\delta}=0$.\\
For any $j=1, \ldots, M$, we extrapolate backward linearly the truncated prediction function to get its value at time $t_{j,-\delta}$, $p_{j,-\delta}$, and set $v_{t_{j,-\delta}}=0$. We assume that the initial transition $(V_{j, 0}|v_{j,-\delta};\bm{\theta},\delta)$ 
has a Beta distribution and apply to it the same moment matching method used above. Given a vector of parameters $\bm{\theta}$, we estimate $\delta$ solving the following problem
\begin{equation}
\arg\max_{\delta}\tilde{\mathcal{L}}_{\delta}\left(\bm{\theta},\delta; v^{M,1}\right) = \arg\max_{\delta}\prod\limits_{j=1}^M \rho_0 \left(v_{j, 0}|v_{j,-\delta};\bm{\theta},\delta\right),
\label{eq:likelihood_delta}
\end{equation}
where $\tilde{\mathcal{L}}_\delta$ is the approximated $\delta-$likelihood. To solve this problem, we repeat the steps described in Subsection \ref{Sec:MinLH}, with the additional initial step of creating the linear extrapolation for $p_{j,-\delta}$ at each $j\in\{1,2,\dots,M\}$.

As anticipated in Remark \ref{remdelta}, we extend the statistical model (\ref{likelihood}) to include the extra parameter $\delta$. The approximated complete likelihood $\tilde{\mathcal{L}}_c$, which estimates the vector $(\theta_0,\alpha,\delta)$, is given by
\begin{equation}
\tilde{\mathcal{L}}_c\left(\bm{\theta},\delta; v^{M,N +1}\right)=\tilde{\mathcal{L}}\left(\bm{\theta}; v^{M,N +1}\right)\tilde{\mathcal{L}}_{\delta}\left(\bm{\theta},\delta; v^{M,1}\right),
\label{eq:complete_LH}
\end{equation}
where $\tilde{\mathcal{L}}\left(\bm{\theta}; v^{M,N +1}\right)$ is the non-log version of (\ref{likelihood}). As we can provide initial guesses for $\bm{\theta}$ and $\delta$, we have a starting point for the numerical optimization of the approximated complete likelihood (\ref{eq:complete_LH}).



\section{Application: the April-December 2019 Uruguay wind and forecast dataset } \label{Section_5}

In recent years, Uruguay has triggered a remarkable change in its energy matrix. In (\cite{irena}, p.23), Uruguay was among those countries showcasing innovation, like Denmark, Ireland, Germany, Portugal, and Spain, with proven feasibility of managing annual variable renewable energy (VRE) higher than 25\% in power systems. 

According to (\cite{ren21}, pp.118--119), in 2018, Uruguay achieved 36\% of its electricity production from variable wind energy and solar PV, raising the share of generation from wind energy more than five-fold in just four years, from 6.2\% in 2014 to 33\% in 2018. 

At present, Uruguay is fostering even higher levels of wind penetration by boosting regional power trading with Argentina and Brazil. 
In this rapidly evolving scenario, it is essential to analyze national data on wind power production with wind power short-term forecasting to orientate and assess the strategies and decisions of wind energy actors and businesses. 

Our study is based on publicly available data (source: Administrator of Electric Market) on the wind power production in Uruguay between April and December 2019, that we adequately normalized with respect to the present \\ $\SI{1474}{\mega\watt}$ maximum installed wind power capacity. Each day, wind power production recordings are available every ten minutes.  In this work, we also considered data from three different forecast providers, available each day starting at 1 pm.

\begin{figure}[H]
\centering
\includegraphics[width=0.485\textwidth]{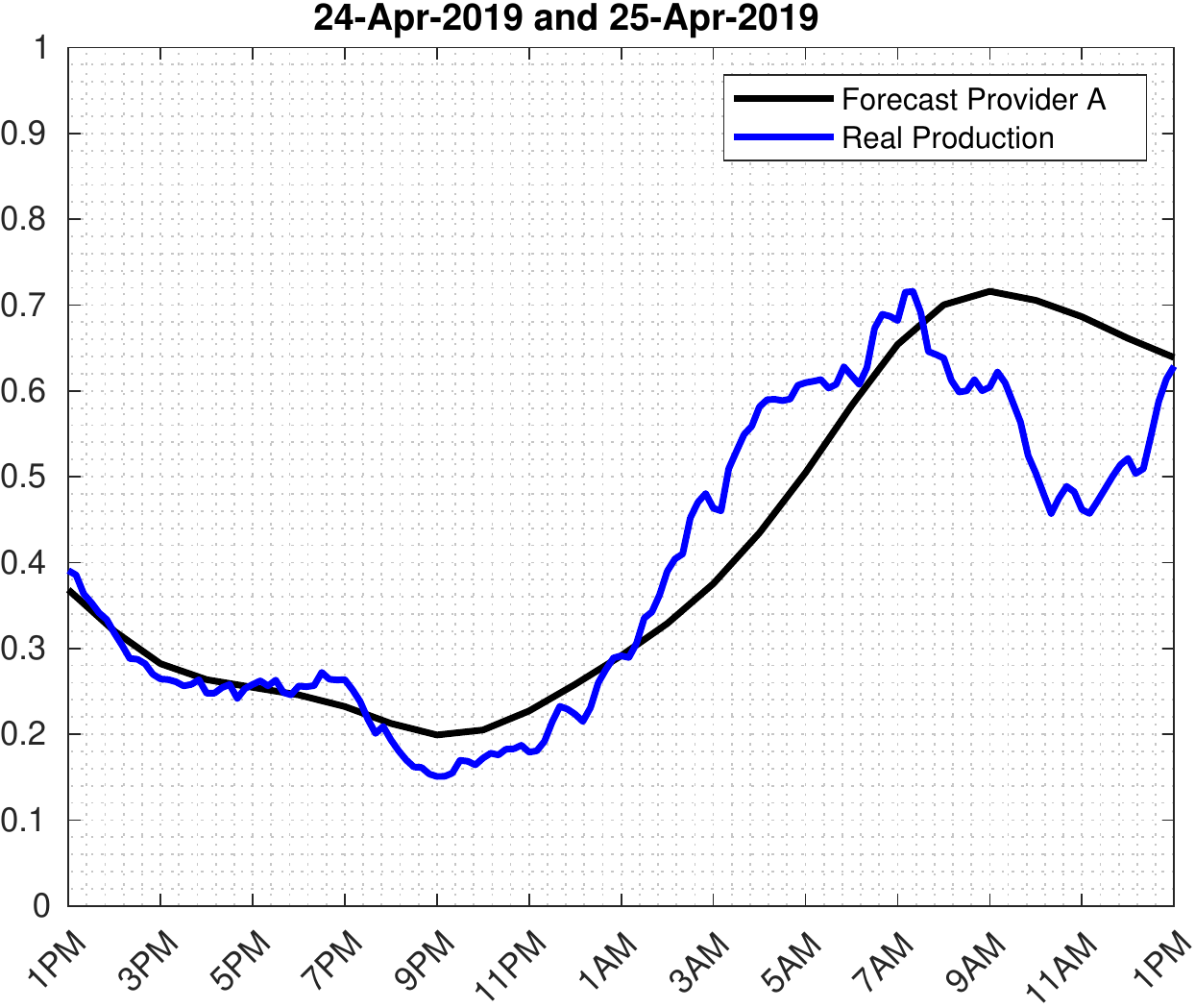}\quad
\includegraphics[width=0.485\textwidth]{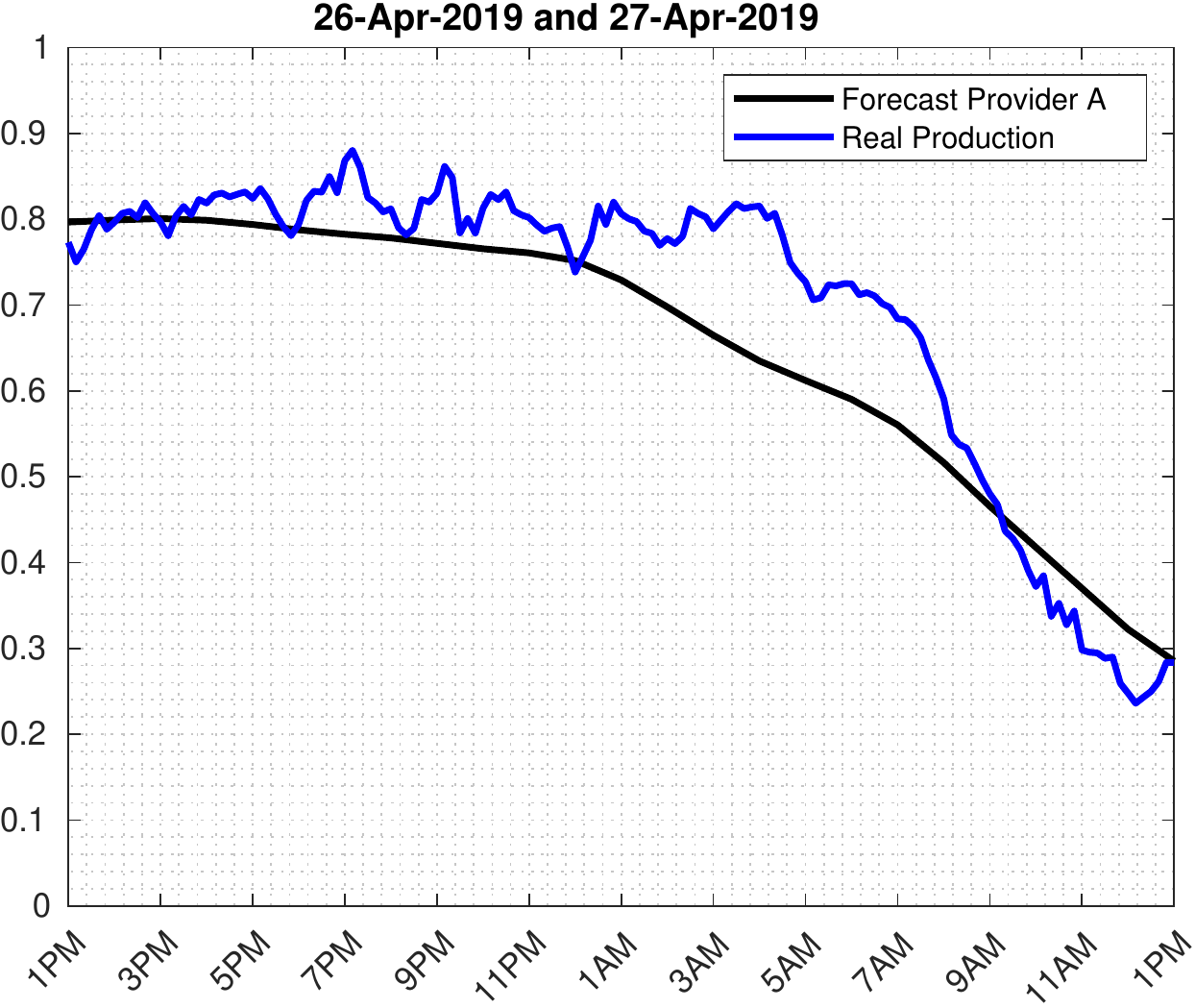}\\
\caption{Two 24-hour segments with the normalized wind power real production in Uruguay (blue line) recorded every ten minutes, and the hourly wind power production forecasted by provider A (black line).}
  \label{fig:sample_data}
\end{figure}

Figure \ref{fig:sample_data} shows the wind power real production during two segments 24 hours selected from the observation period together with their corresponding hourly short-term forecast, computed by a forecast provider. For the sake of visualization clarity, this section relies only on forecasts from one provider, called ``provider A'' from now on, ranked as the most accurate forecast provider, as it emerged from our posterior analysis.

A view of the global discrepancy between the real production and the forecasted production, during the nine months observation period, is summarized through the forecast error histograms in the next Figure \ref{fig:data_curtailing}, where we also partitioned the forecast errors according to three contiguous categories of normalized generated power. Low normalized generated power corresponds to the range $[0,0.3]$, mid-power refers to the range $]0.3,0.6]$, and high-power to the range $]0.6,1]$.

\begin{figure*}
\centering
\includegraphics[width=0.485\textwidth]{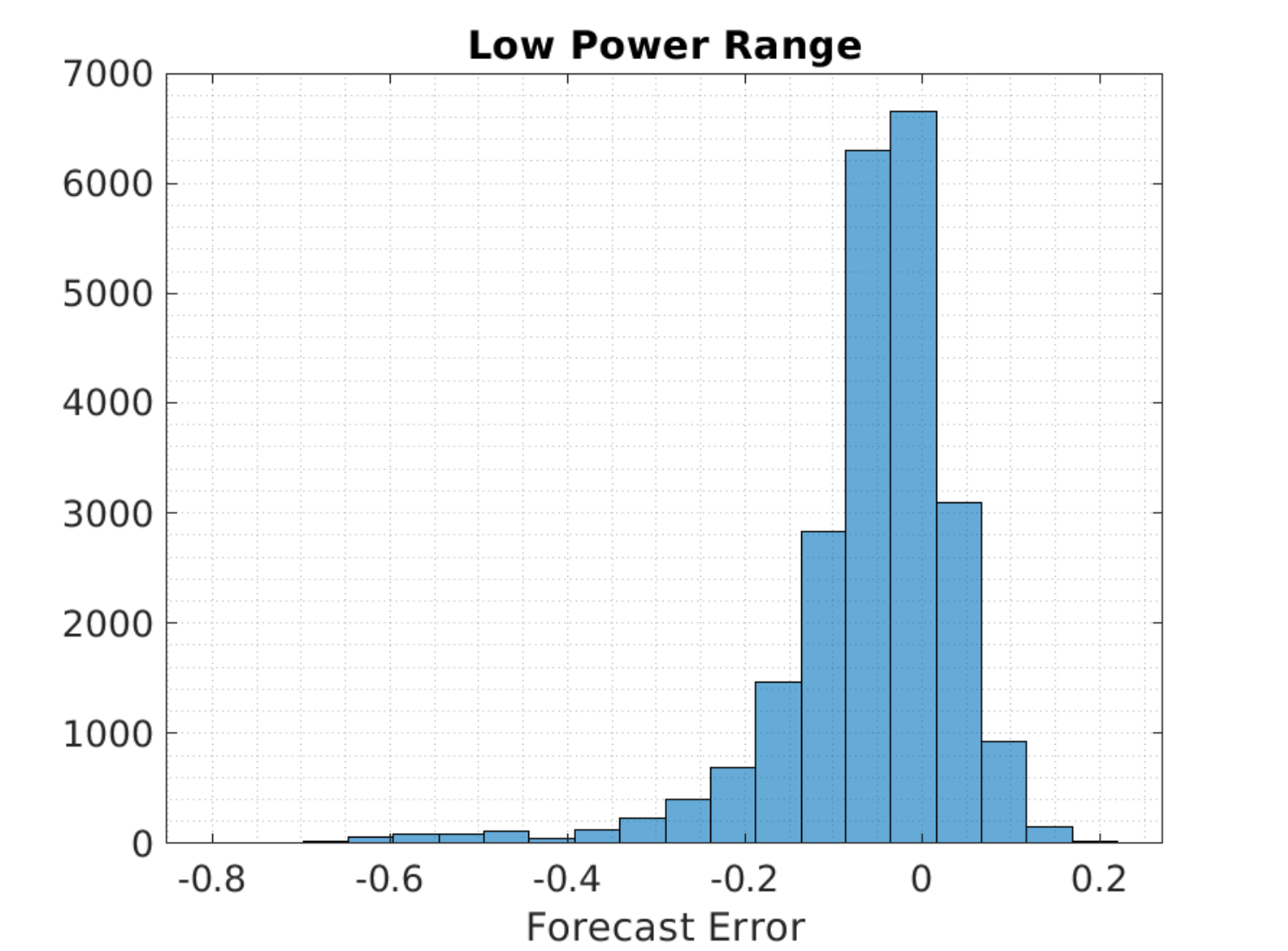}
\includegraphics[width=0.485\textwidth]{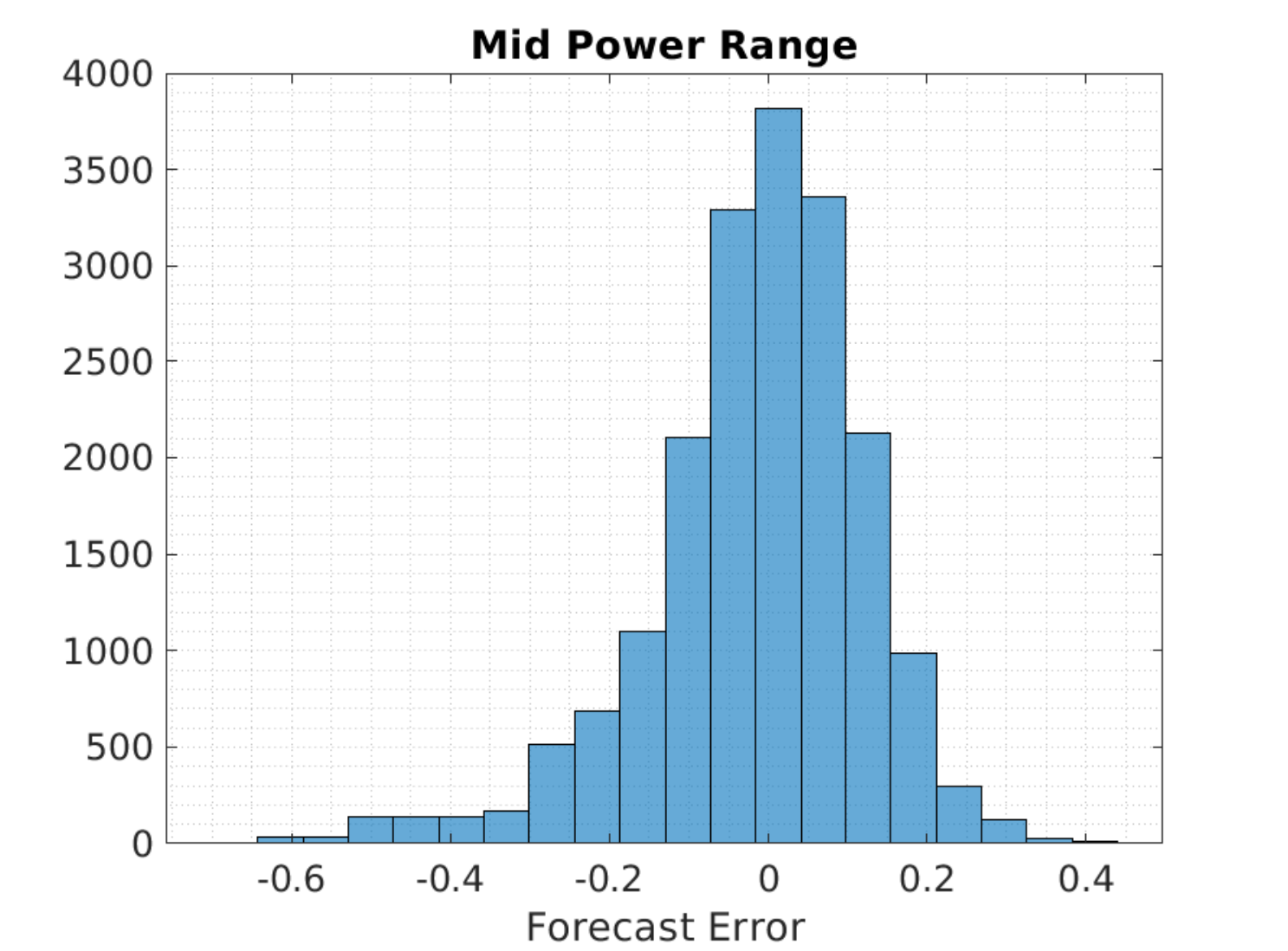}\\
\quad\\
\includegraphics[width=0.485\textwidth]{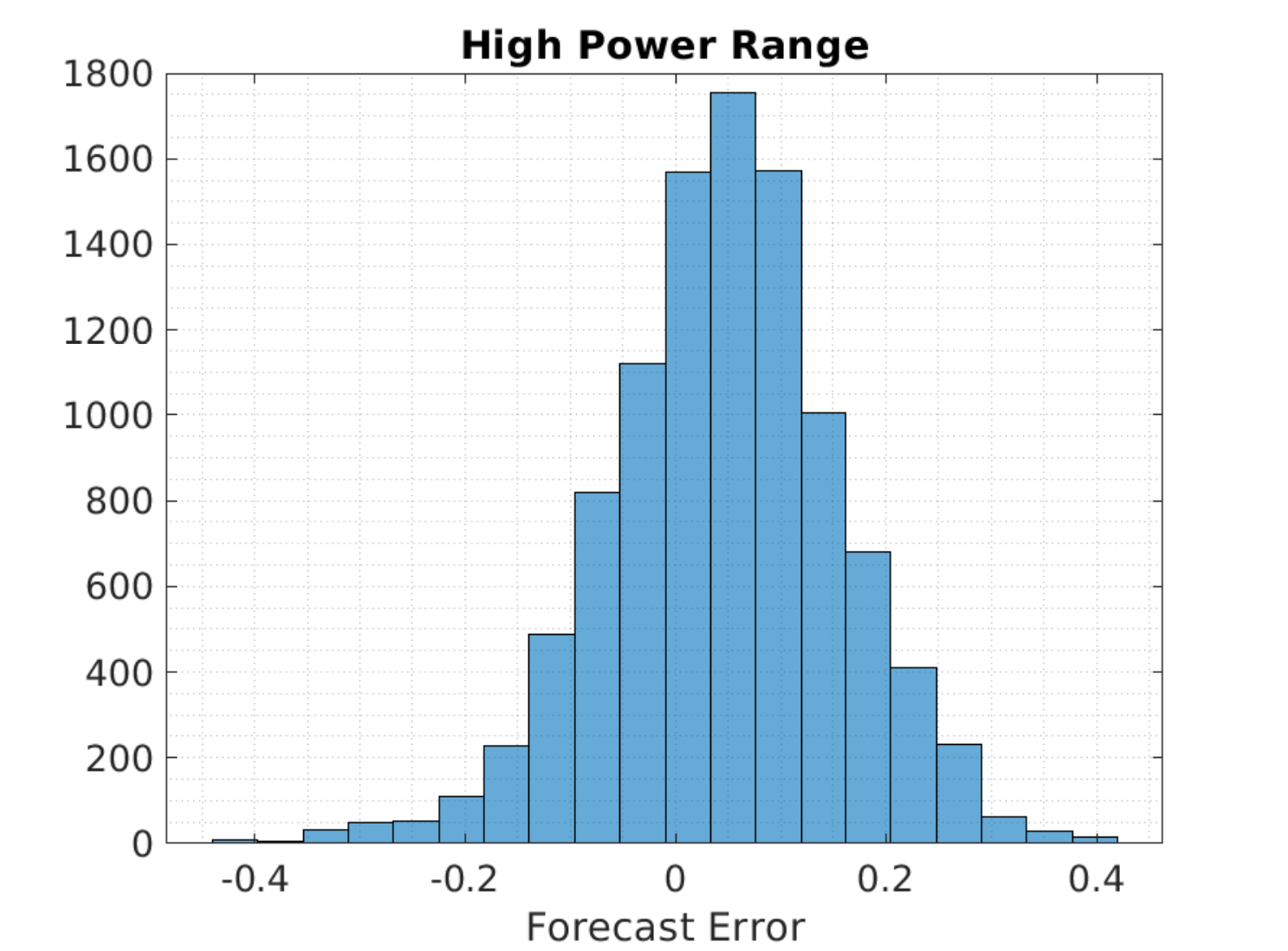}
\includegraphics[width=0.485\textwidth]{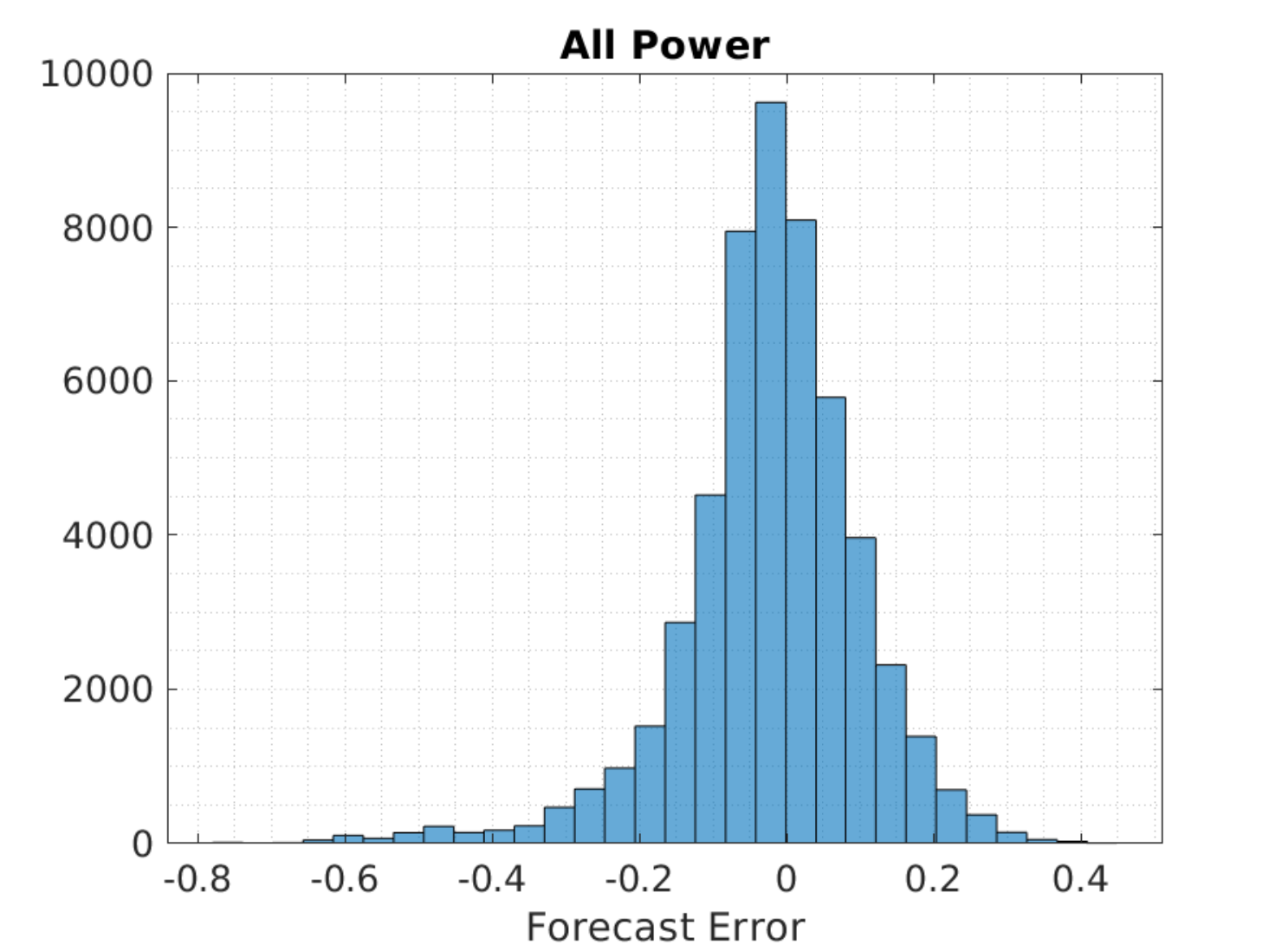}
\caption{Wind production forecast error histograms during the period April-December 2019: low-power (upper-left plot), mid-power (upper-right plot), high-power (lower-left plot), and the global range of power (lower-right plot).}
  \label{fig:data_curtailing}
\end{figure*}

We may observe that all the histograms in Figure \ref{fig:data_curtailing} exhibit skewed patterns, to a different extent, as well as extreme observations. The presence of these features can be partly explained. The data analysis highlighted that, during several 24-hour segments, the system operators decided to reduce or even cease the wind power production. Indeed, as recalled in (\cite{irena2}, p.8), ``Uruguay experiences high curtailment levels because generation exceeds demand.'' Despite the large country's interconnection capacity with Argentina and Brazil, there is no active cross-border market; the energy is traded via ad hoc short-term agreements. (\cite{irena2}, p.3) ``Even with interconnection capacity exceeding peak demand, the power system experiences high VRE curtailment, mostly at night when wind generation exceeds demand.''
  
The curtailment of the wind power production imposed by the system operators has a strong influence on the forecast error. To build a model that, driven by the available forecast, allows the inclusion of true power production with a prescribed degree of uncertainty, it is necessary to remove the data segments affected by wind curtailment.

Once we removed all the 24-hour segments showing wind curtailment, we set up a dataset containing 147 daily segments. 
In the absence of the curtailment intervention, the forecast error histograms shown below in Figure \ref{fig:data_after_clean}, can appreciate skewness reduction, except for low power forecast error histogram.

\begin{figure*}
\centering
\includegraphics[width=0.485\textwidth]{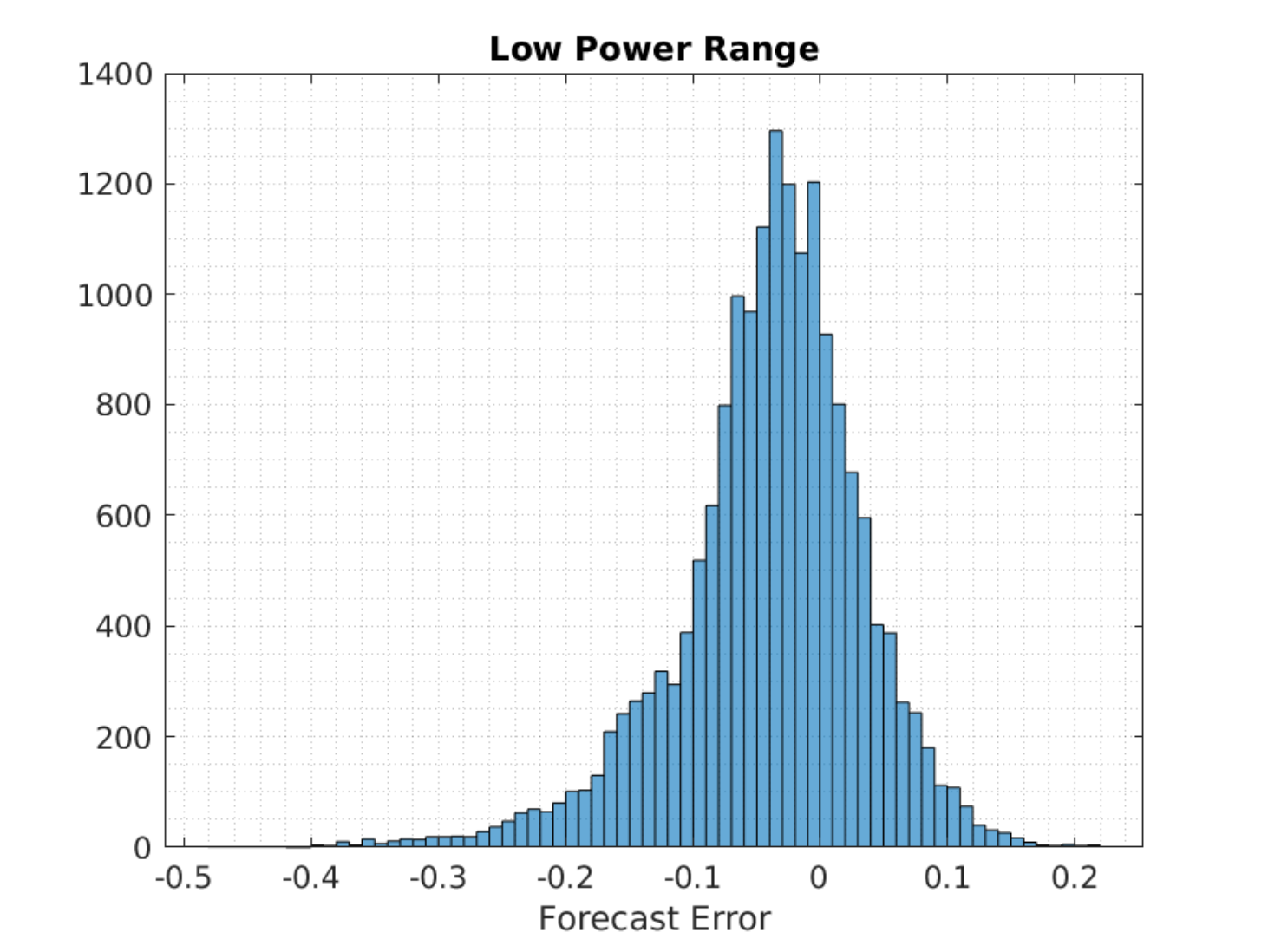}
\includegraphics[width=0.485\textwidth]{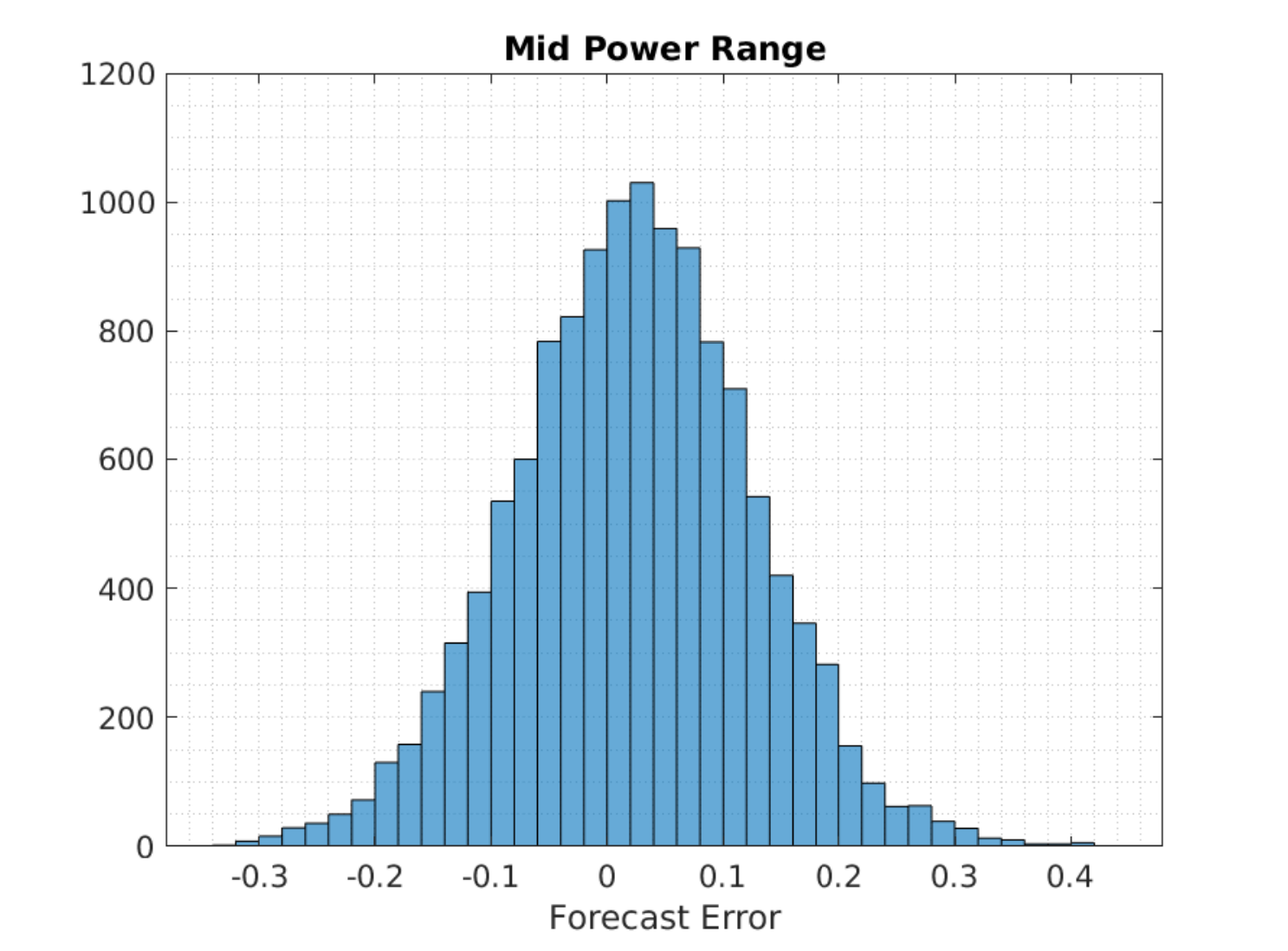}\\
\quad\\
\includegraphics[width=0.485\textwidth]{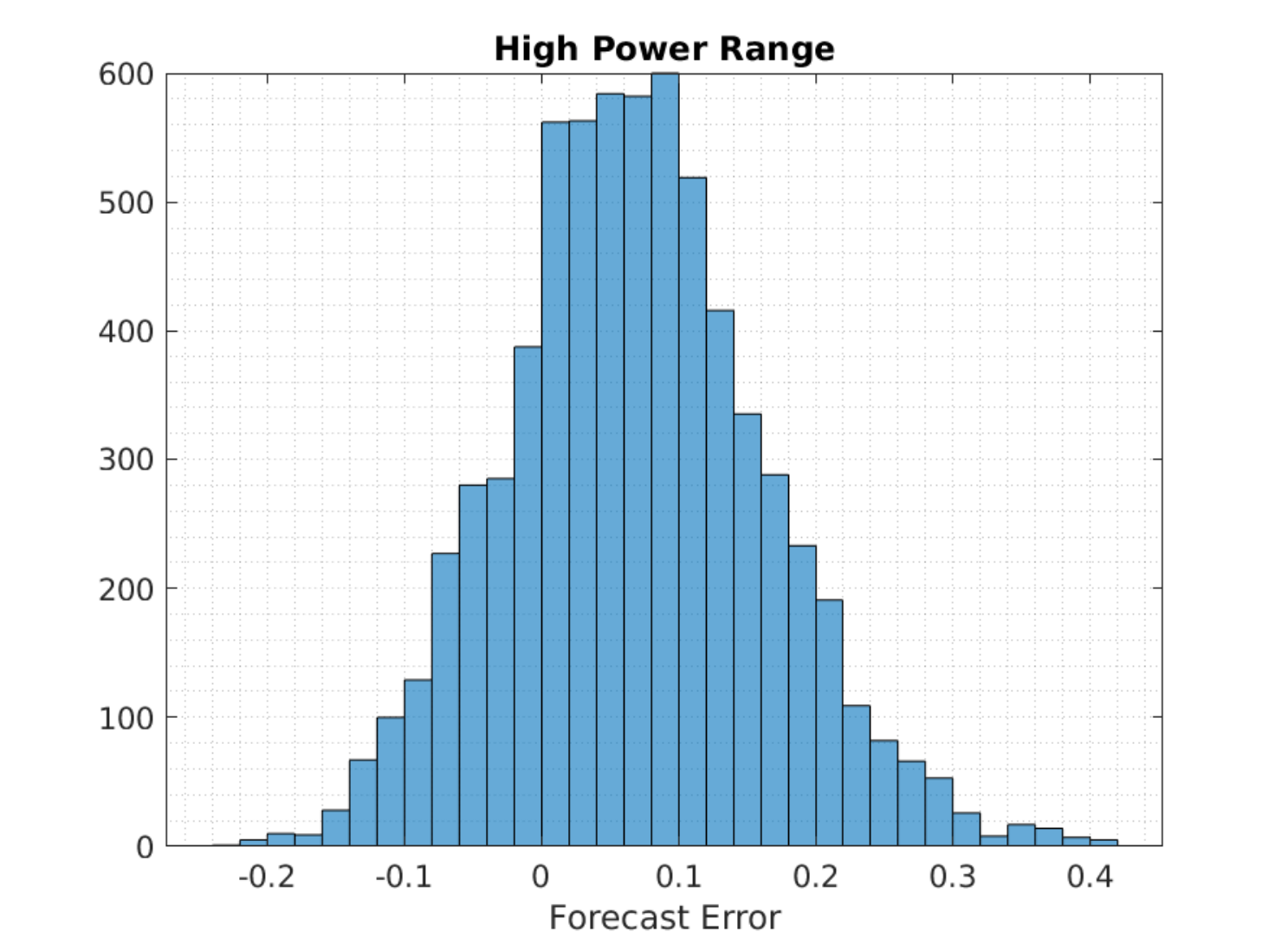}
\includegraphics[width=0.485\textwidth]{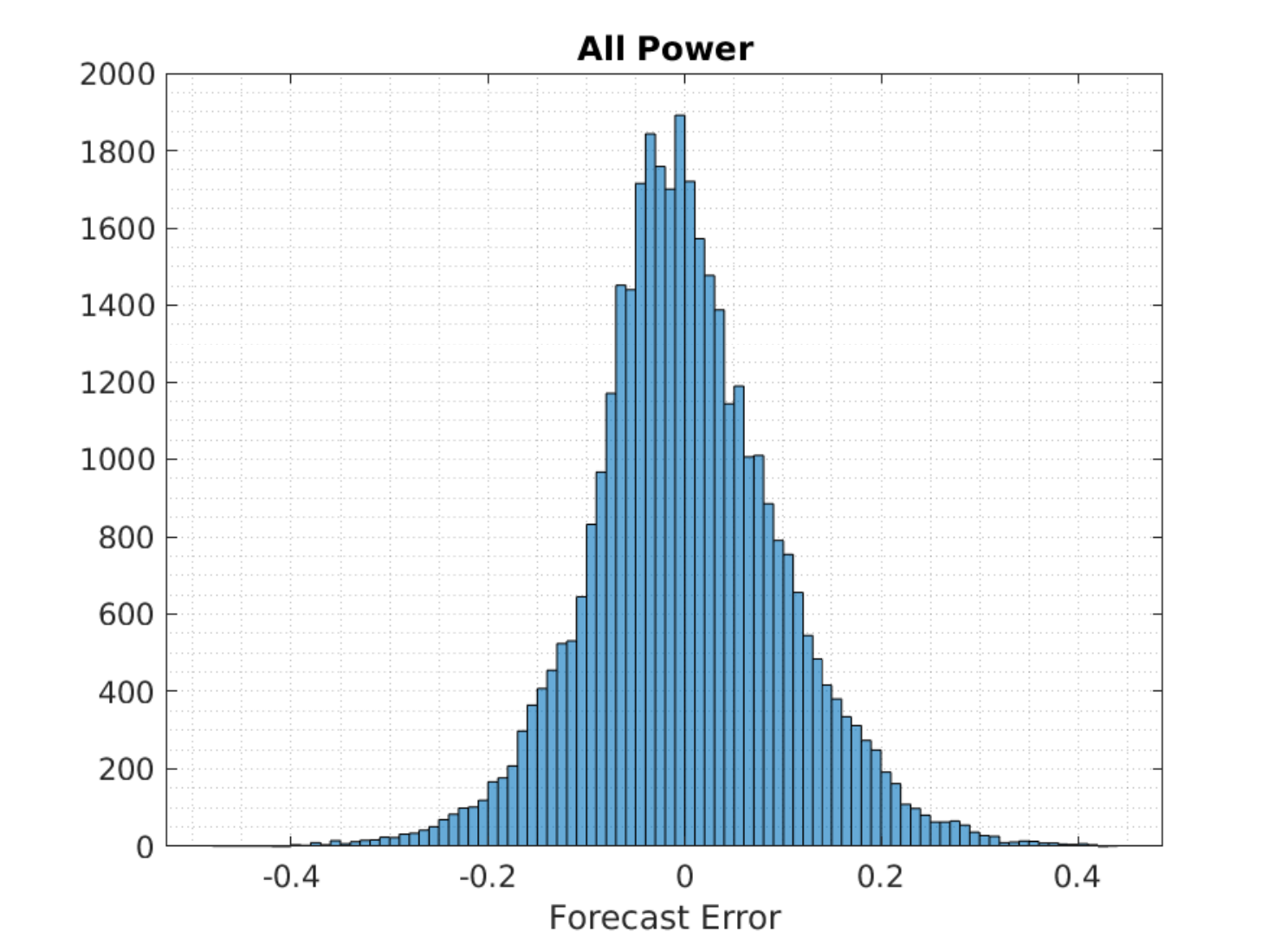}
\caption{Wind production forecast error histograms during the period April-December 2019 after removing 24-hour segments with artificial wind curtailment: low-power (upper-left plot), mid-power (upper-right plot), high-power (lower-left plot), and the global range of power (lower-right plot).}
  \label{fig:data_after_clean}
\end{figure*}

In this stage of data preprocessing, we obtain another useful result by applying the first-order difference operator to the forecast errors. 
The forecast error transition histograms, displayed in the next Figure \ref{fig:error_transitions}, will later constitute a reference for the visual assessment of the global fit of the proposed models.

\begin{figure*}
\centering
\includegraphics[width=0.485\textwidth]{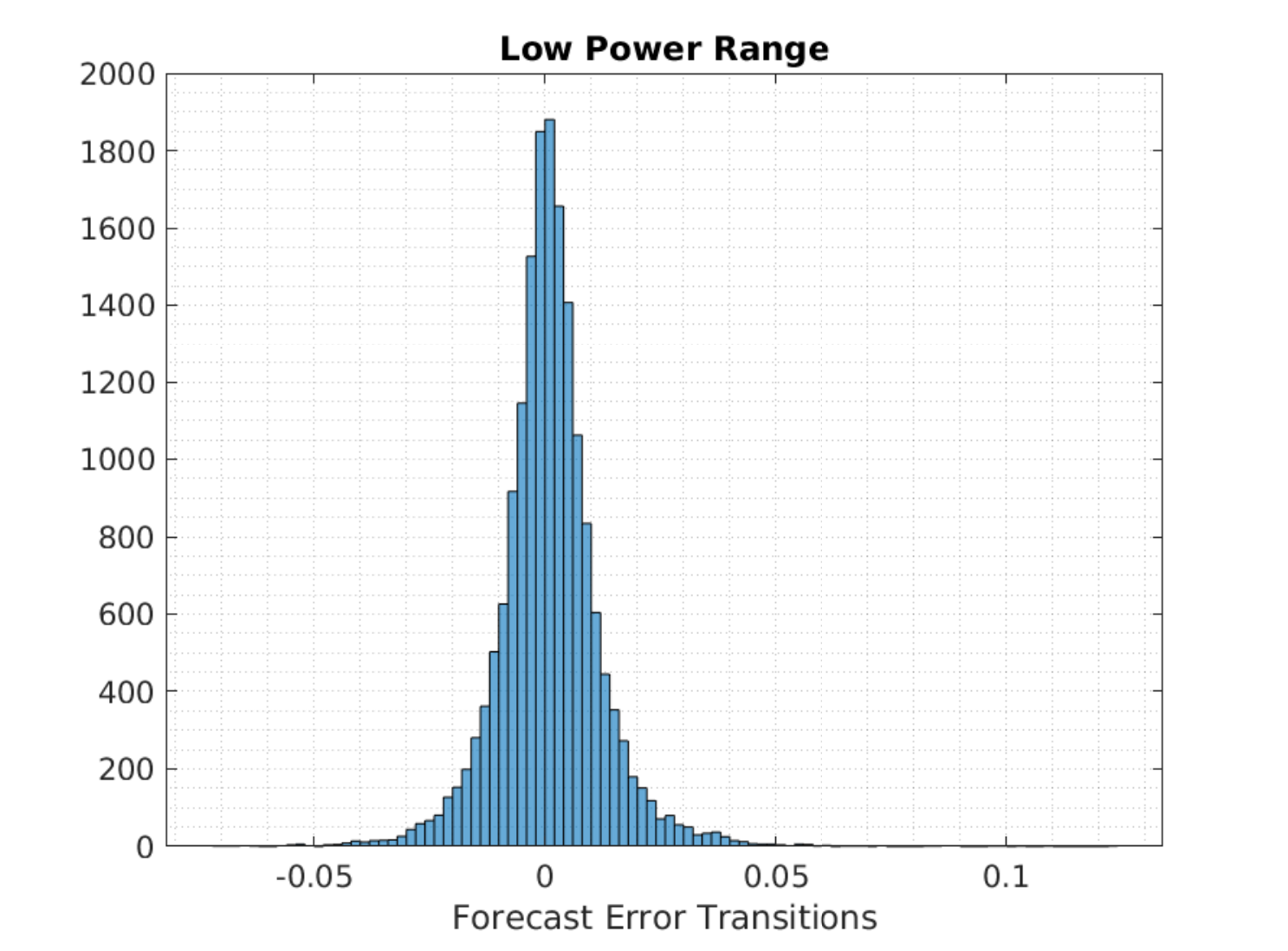}
\includegraphics[width=0.485\textwidth]{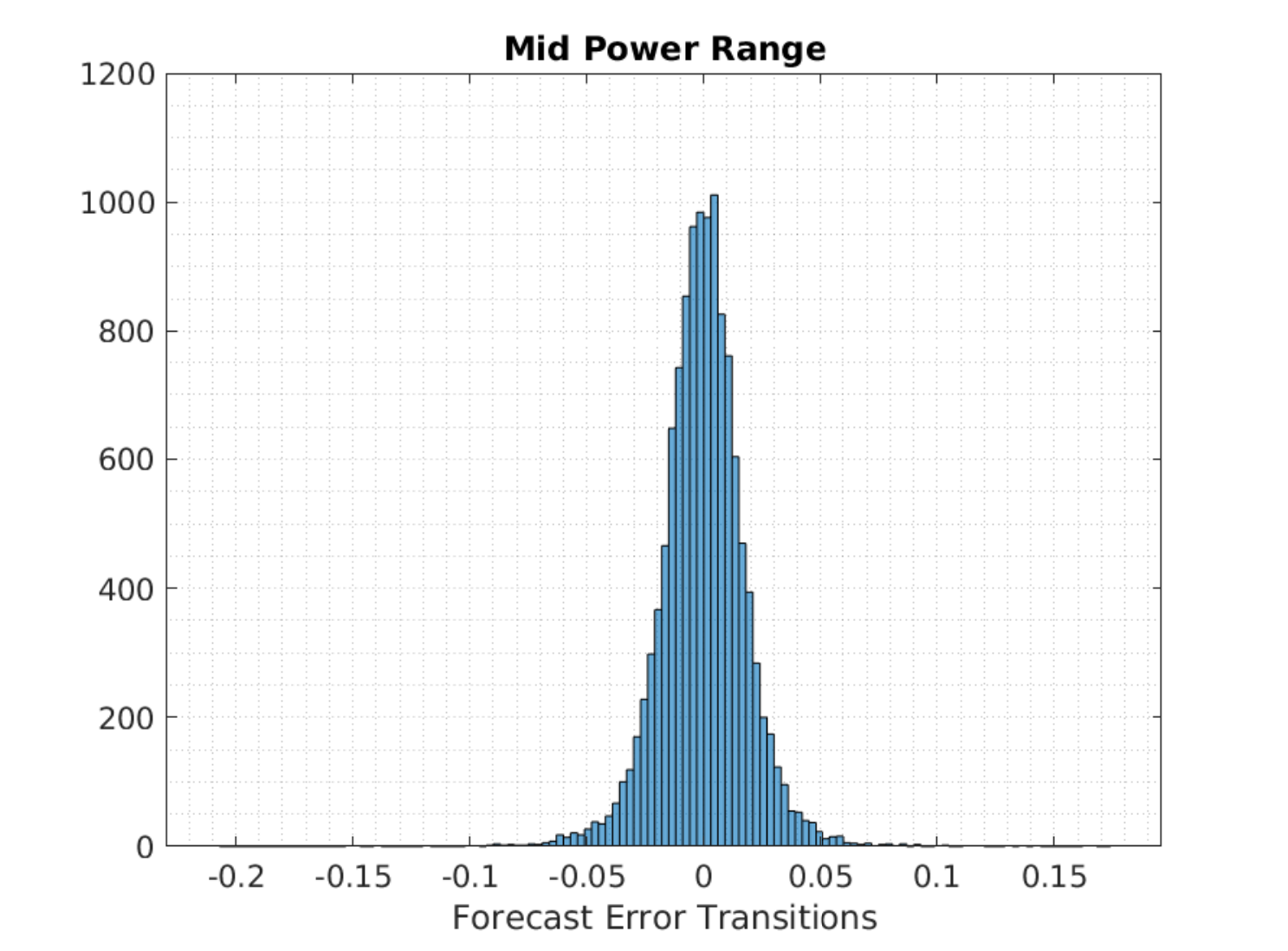}\\
\quad\\
\includegraphics[width=0.485\textwidth]{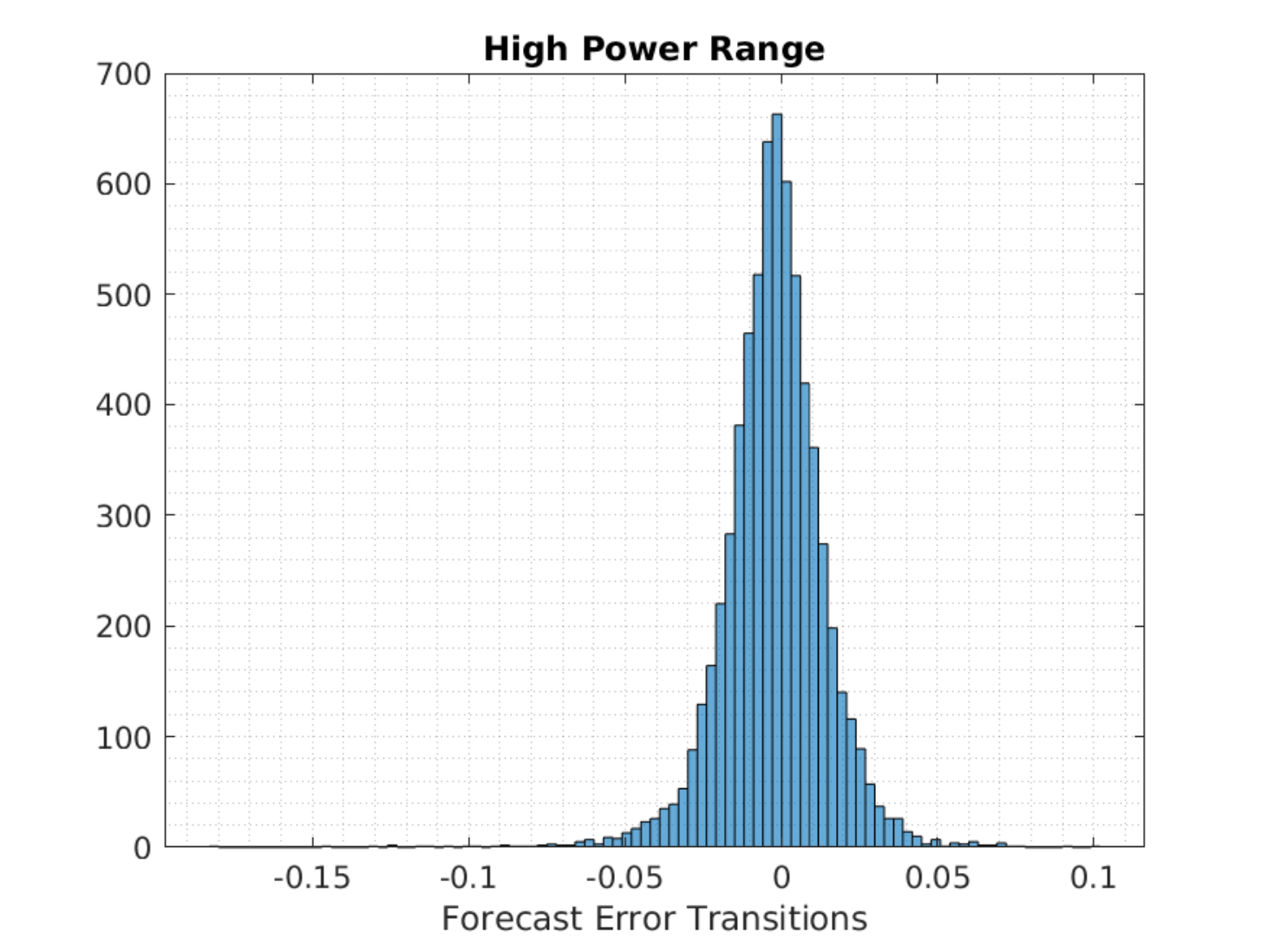}
\includegraphics[width=0.485\textwidth]{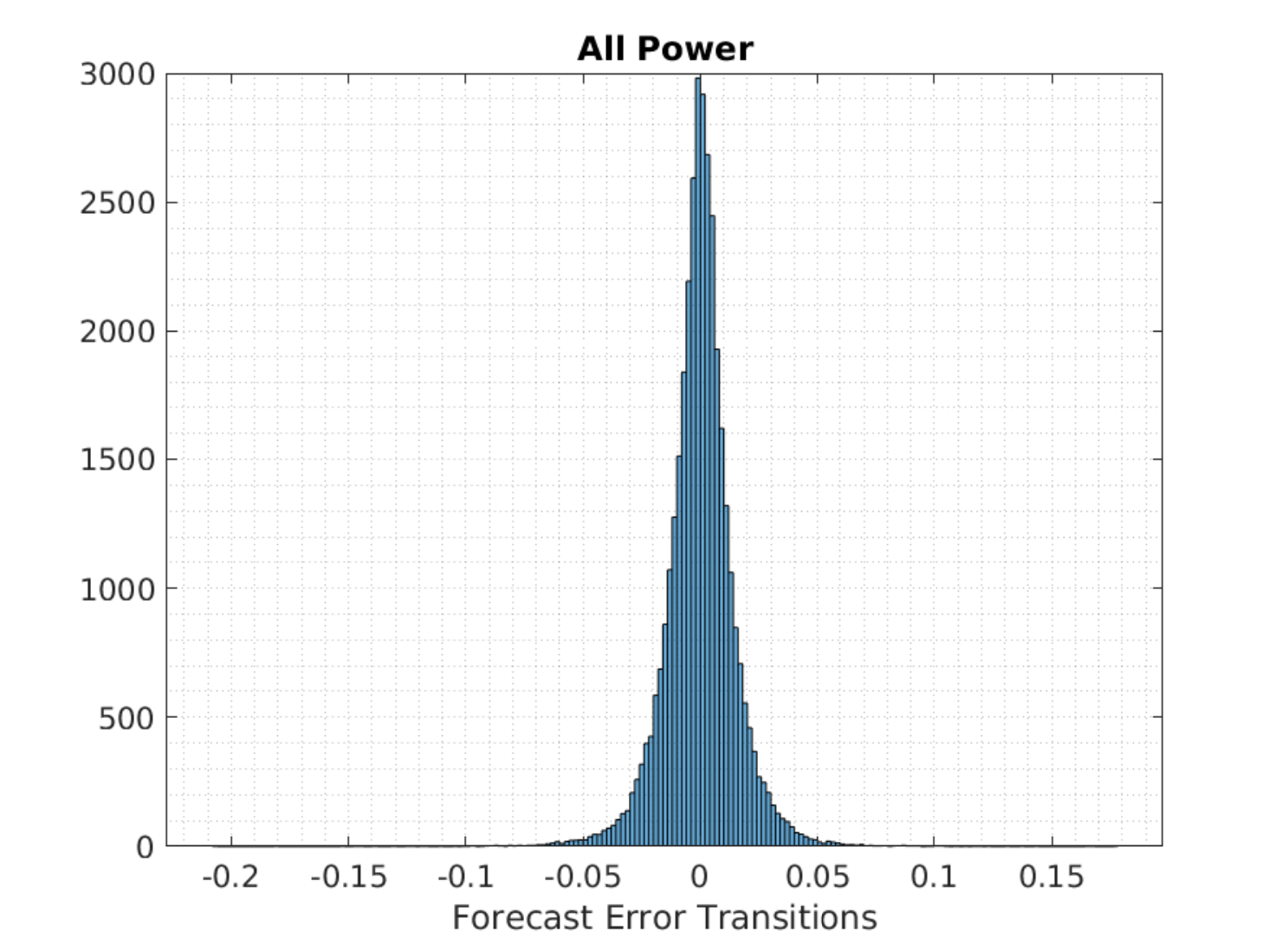}
\caption{Forecast error transition histograms during the period April-December 2019 without wind power production curtailment: low-power (upper-left plot), mid-power (upper-right plot), high-power (lower-left plot), and the global range of power (lower-right plot).}
\label{fig:error_transitions}
\end{figure*}

The histograms in Figure \ref{fig:error_transitions} feature a non-Gaussianity trait and provide initial input for the model-building stage.

\begin{figure*}
\centering
\includegraphics[width=0.485\textwidth]{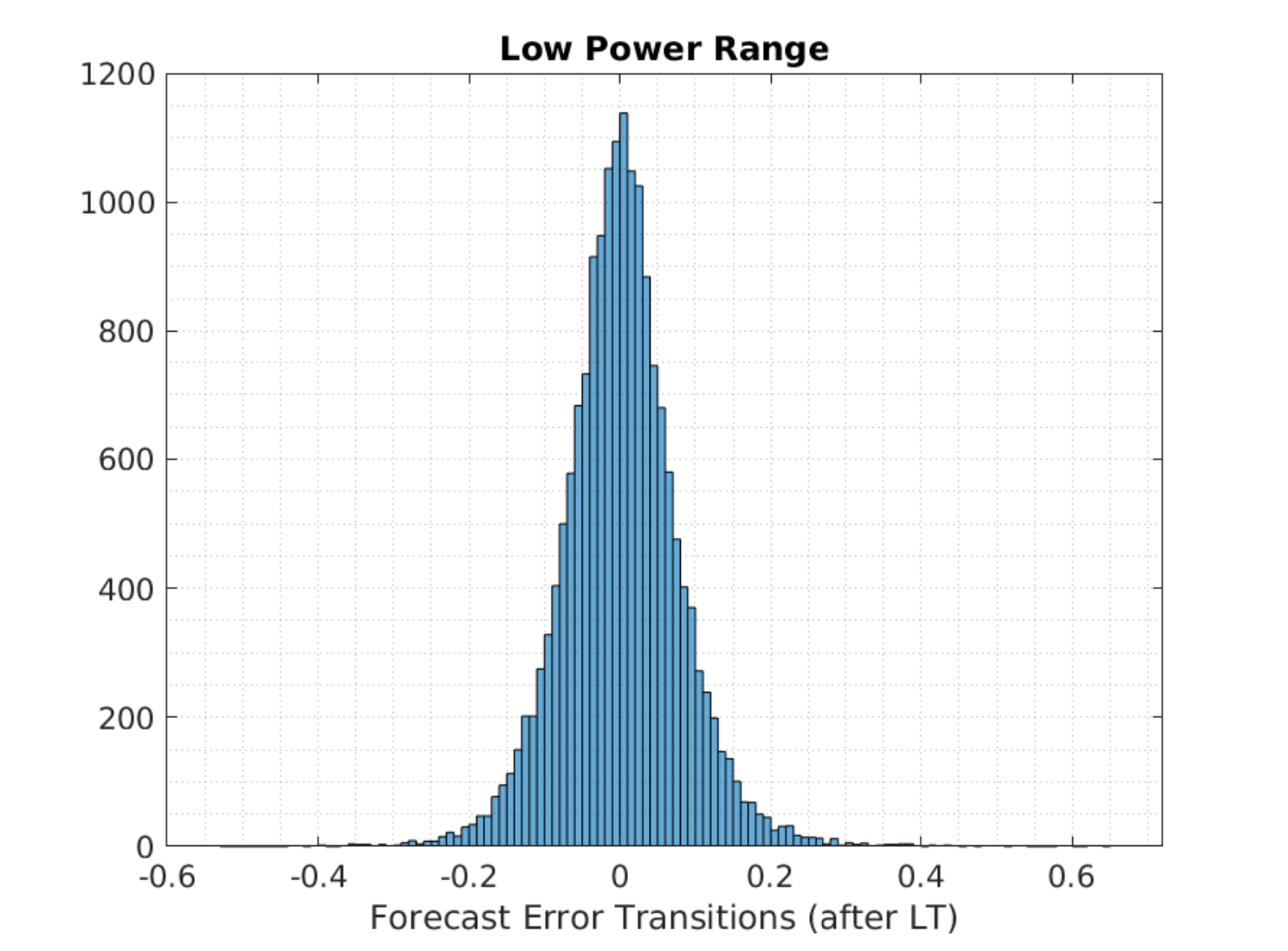}
\includegraphics[width=0.485\textwidth]{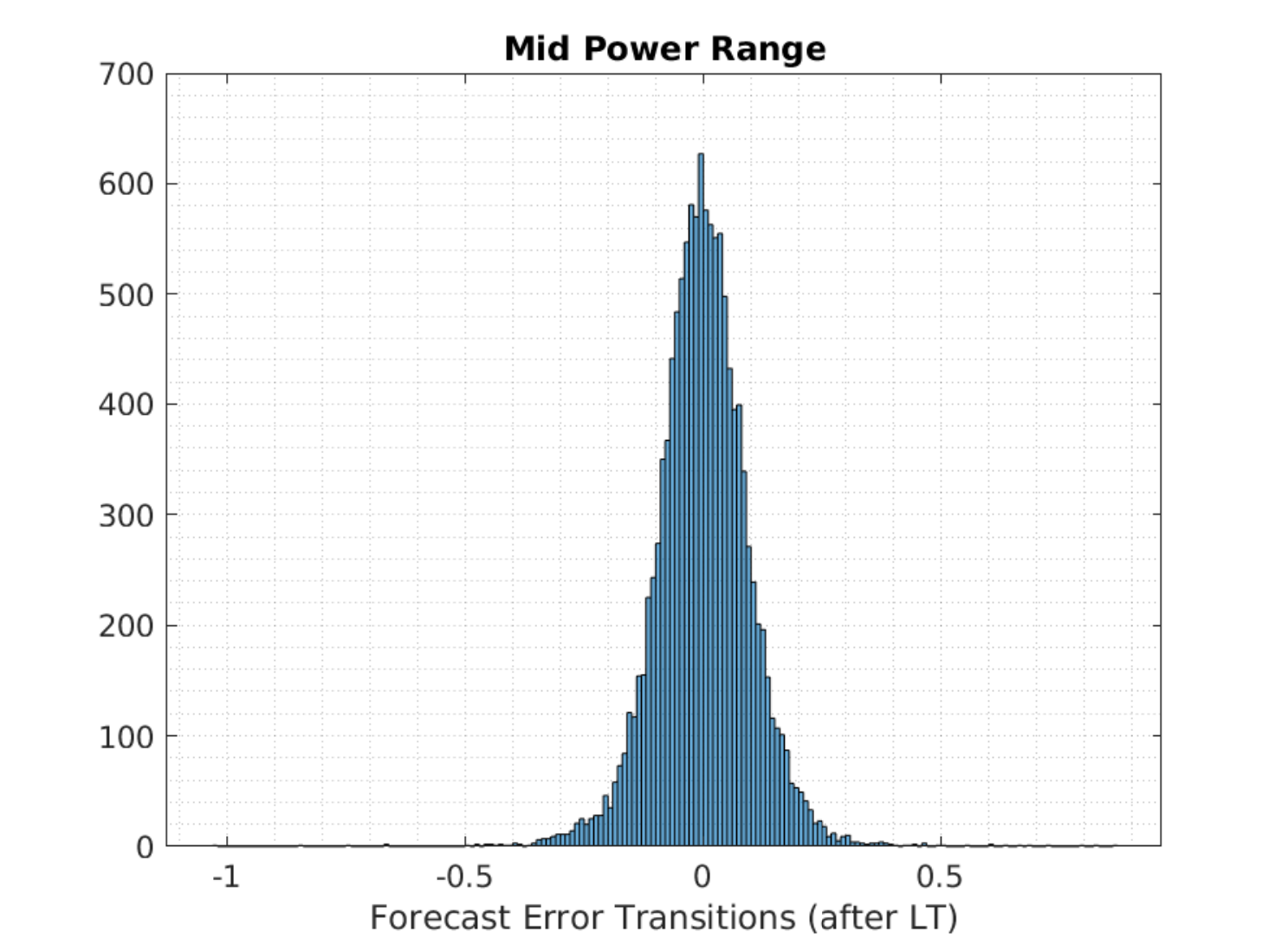}\\
\quad\\
\includegraphics[width=0.485\textwidth]{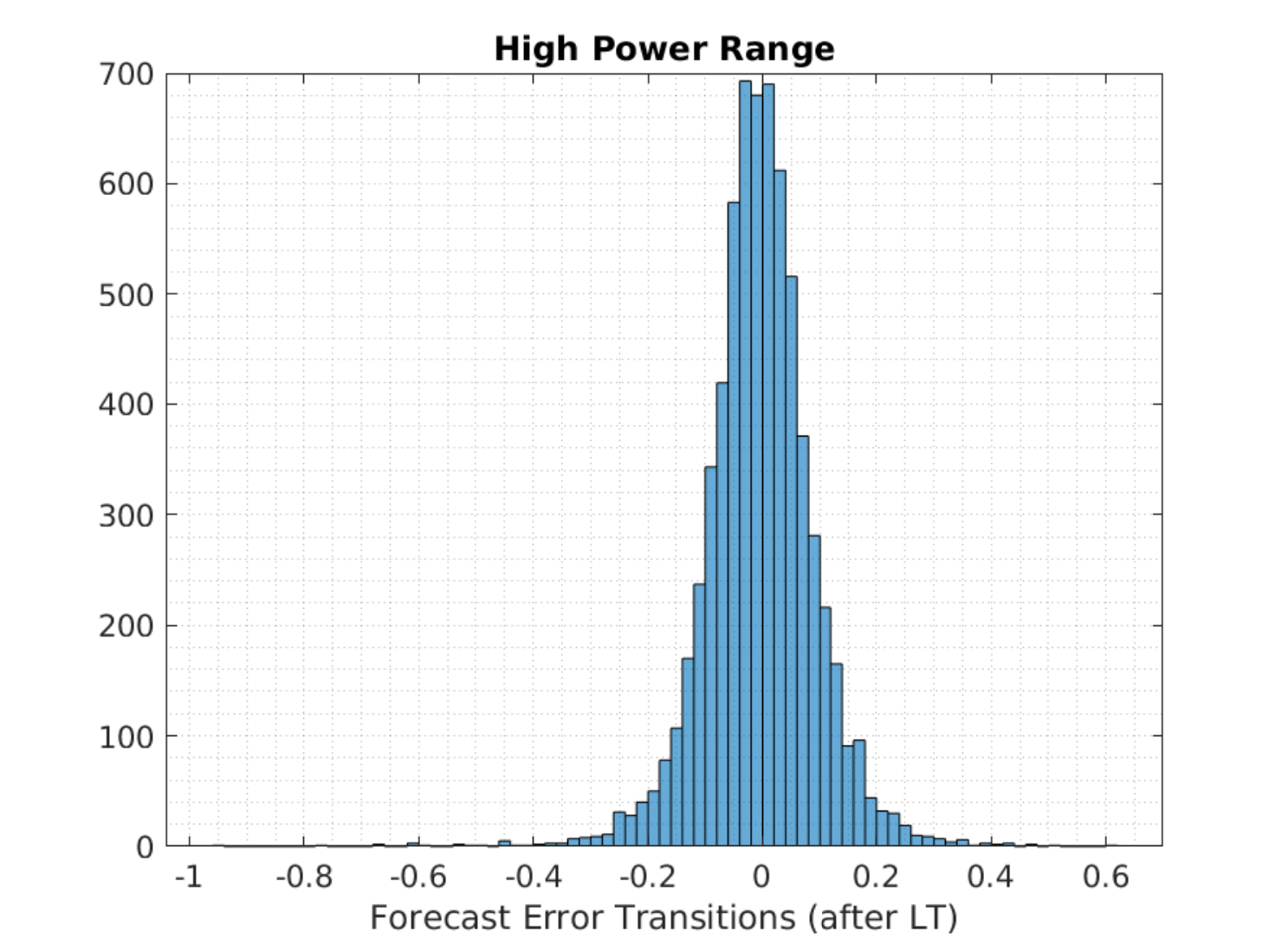}
\includegraphics[width=0.485\textwidth]{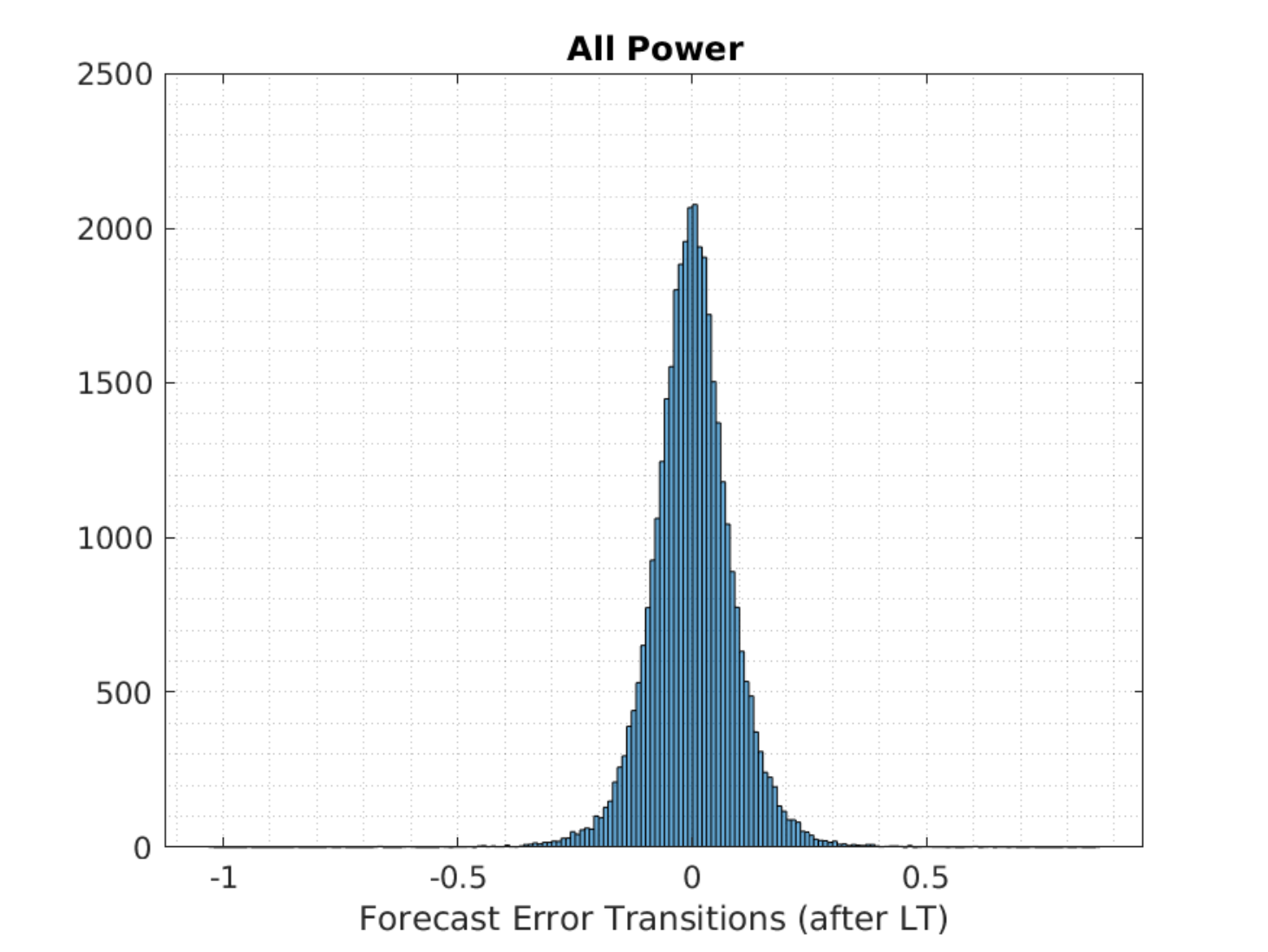}
\caption{Lamperti transformed forecast error transition histograms between April and December 2019 without wind power production curtailment: low-power (upper-left plot), mid-power (upper-right plot), high-power (lower-left plot), and the global range of power (lower-right plot).}
  \label{fig:LP_transitions}
\end{figure*}

Guided from inferring the unknown model parameters, we also propose transforming data as a strategy that naturally leads to an alternative model. \\
In this case, the Lamperti transform has been applied using the optimal estimates of the parameters in the SDE model (\ref{VtSDE}), obtained applying our numerical procedure detailed in Subsection \ref{opt_sec}. See the histograms in Figure \ref{fig:LP_transitions}.

\subsection{Calibration of the approximate negative log-likelihood in $V$-space and $Z$-space}

To implement the models' calibration procedure, we select 73 non-contiguous segments of normalized wind power production out of the 147 segments, each 24-hours long, assigning them to the training set.
The other 74 non-contiguous segments compose the test set. Such an allocation mechanism guarantees independence among the segments, matching the assumption we did in Section \ref{Section_4} to formulate the statistical models. {Additional cross-correlation tests were performed to ensure this assumption.}

All the following results involving a single provider refer to provider A. Furthermore, all calibrations involve the training sets and all simulations, the test sets. Following the instruction for the initial guesses from Subsection $\ref{opt_sec}$ and assuming that
\begin{equation}
\theta_t=\max\left(\theta_0,\frac{\alpha\theta_0+|\dot{p}_t|}{\min(p_t,1-p_t)}\right),
\label{eq:applied_thetat}
\end{equation}
we obtain the initial guess $(\theta_0^*,\alpha^*,\delta^*)\approx(1.54,0.072,073)$.

As an auxiliary verification, we plot the negative log-likelihood (negative version of (\ref{eq:loglikelihoodV})) as a function of the parameters, and we use additional minimization functions from MATLAB R2019b. {Moreover, we realized an additional inference utilizing the test sets to guarantee the robustness of our numerical methods.}

\begin{figure}[H]
\centering
\includegraphics[width=0.485\textwidth]{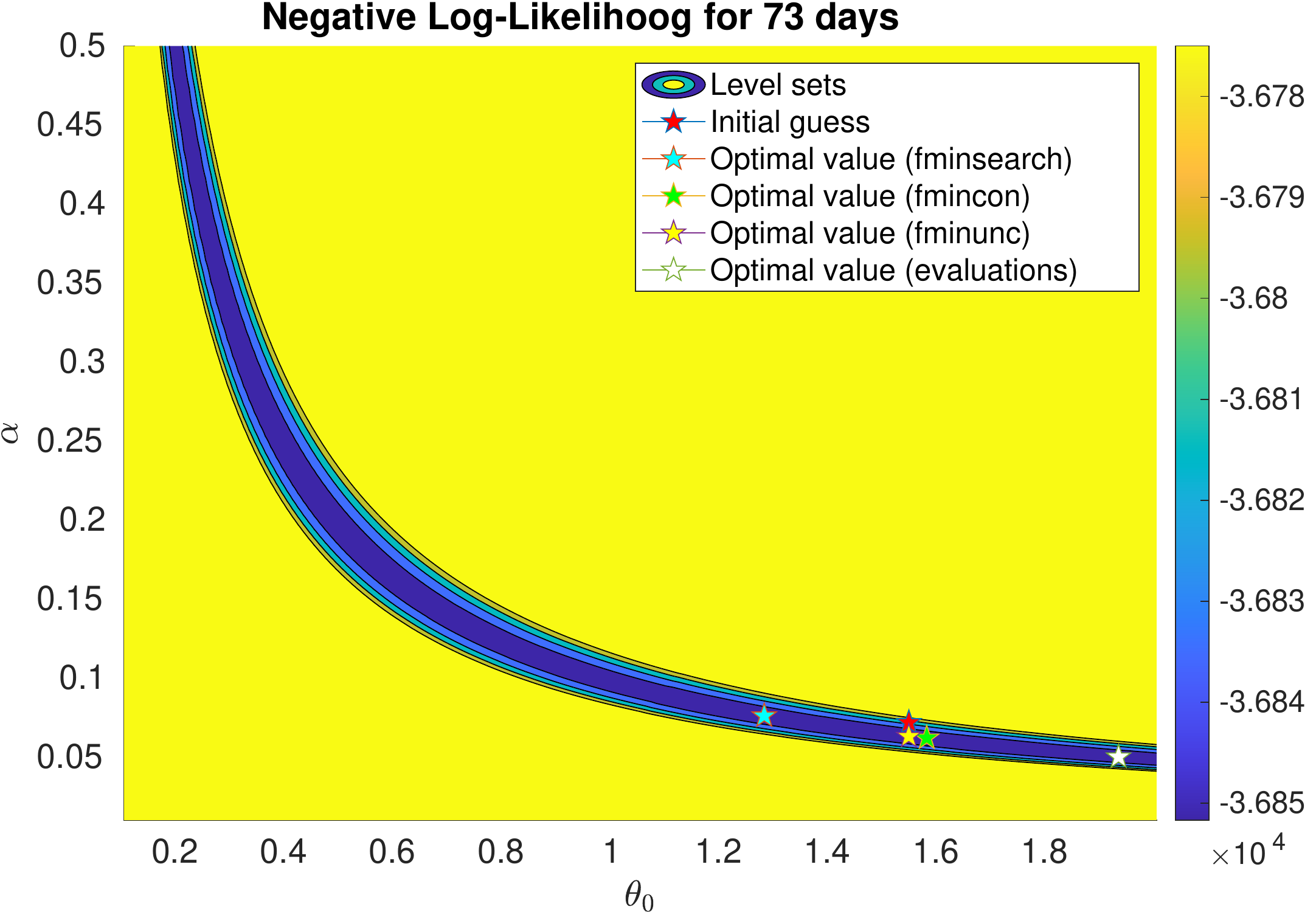}\quad\includegraphics[width=0.485\textwidth]{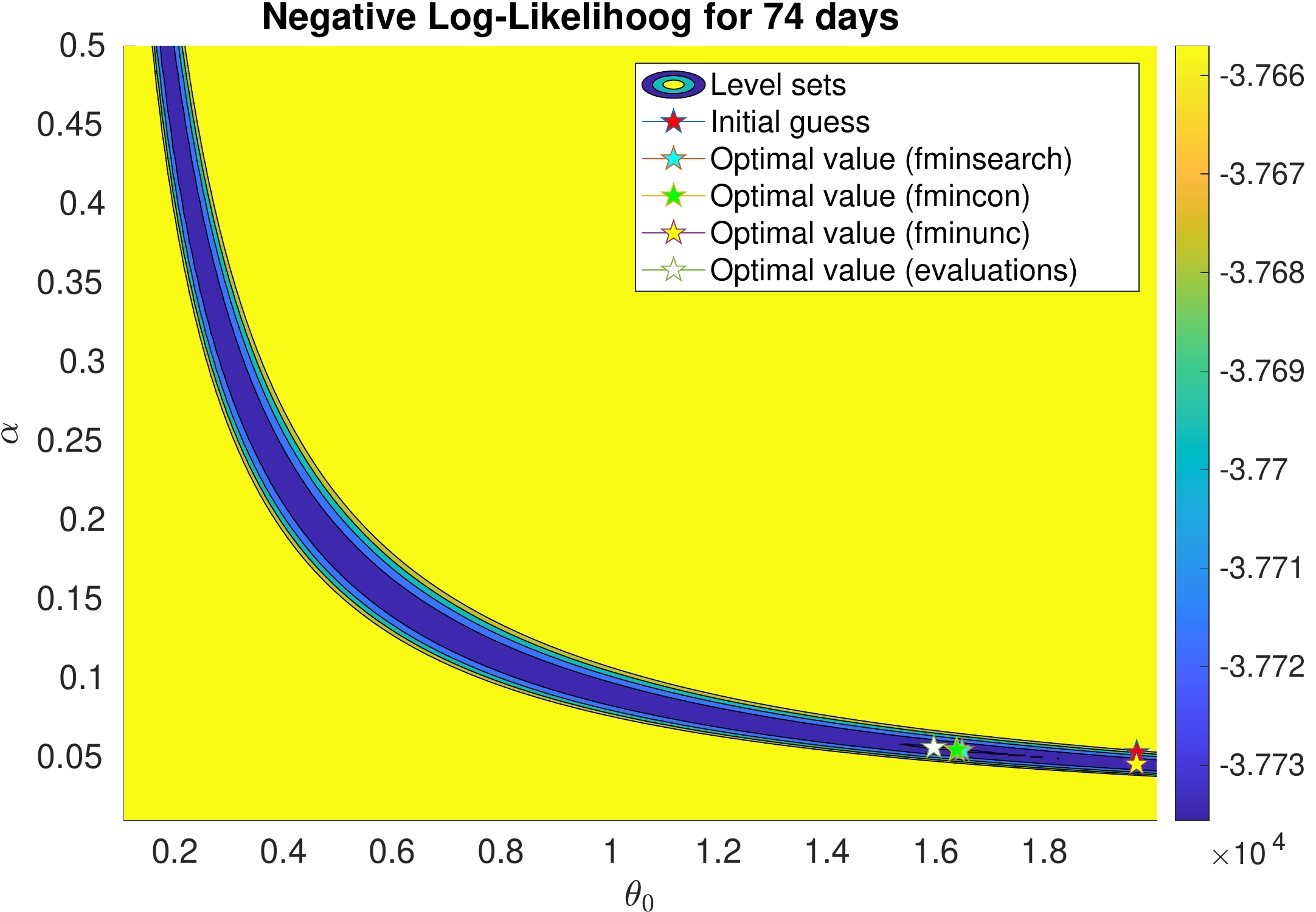}
\caption{{Negative log-likelihood's level sets for the training sets (plot on the left), and for the test sets (plot on the right). All optimal values are located over the curve $\theta_0\alpha=0.097$ and $\theta_0\alpha=0.089$ for the training and test sets, respectively.}}
\label{fig:neg-LL}
\end{figure}
{On Figure (\ref{fig:neg-LL}), we can see the level sets for the negative log-likelihood for both training and test sets.} The numerical values of each relevant point can be seen in Table (\ref{tab:optimal_values}). We set the optimal parameters in the $V$-space $(\theta_0^V,\alpha^V)=(1.93,0.050)$, as it is where the negative log-likelihood for the training sets reaches its minimum value.
\begingroup
\renewcommand{\arraystretch}{0.8}
\begin{table}[h]
\caption{Coordinates of the initial guess points and optimum points in Figure \ref{fig:neg-LL}.}
\label{tab:optimal_values}
\centering
\begin{tabular}{lcccccc}
\hline\noalign{\smallskip}
 & \multicolumn{3}{c}{Training sets} & \multicolumn{3}{c}{Test sets} \\
\hline\noalign{\smallskip}
 & $\theta_0$ & $\alpha$ & $\theta_0\alpha$ & $\theta_0$ & $\alpha$ & $\theta_0\alpha$\\
\noalign{\smallskip}\hline \noalign{\smallskip}
 Initial guess & 1.54 & 0.072 & 0.111 & 1.96 & 0.053 & 0.104 \\
 fminsearch & 1.14 & 0.076 & 0.097 & 1.64 & 0.054 & 0.089 \\
 fmincon & 1.58 & 0.062 & 0.097 & 1.63 & 0.055 & 0.089 \\
 fminunc & 1.54 & 0.063 & 0.097 & 1.96 & 0.045 & 0.089 \\
 Evaluations & 1.93 & 0.050 & 0.097 & 1.59 & 0.056 & 0.089 \\
\noalign{\smallskip}\hline
\end{tabular}
\end{table}
\endgroup

We observe that all the local (possibly global) minimizers are located over the curves $\theta_0\alpha=0.097$ and $\theta_0\alpha=0.089$ for the training and test sets, respectively. This effect shows that the optimization is more sensitive to the diffusion than to the drift.

In the $Z$-space, we obtain the optimal parameters $(\theta_0^Z,\alpha^Z)$ $=(1.87,0.043)$. 

\begin{algorithm}[H]
\caption{Fixed-point likelihood optimization approach in the $Z-$space} \label{alg1}
\begin{algorithmic}[1]
\State \textbf{load} the training set with normalized wind power production and forecast data
\State \textbf{compute}  $\theta_t$ as in (\ref{eq:applied_thetat}) for any given point $ \bm{\theta}^{\star }$
\State  \textbf{compute} the Lamperti transform $\{h( v_{j,i},t_{j,i};\bm{\theta}^\star )\}_{j=1,i=0}^{M,N}$ as in (\ref{eq:LampZ})
\State \textbf{apply} the moment-matching technique by solving numerically the initial-value problem (\ref{eq:Ztmom})
\State \textbf{compute} the approximate Lamperti log-likelihood (\ref{loglikelihoodZ})
\State \textbf{compute} $
\arg\max_{\bm{\theta}}\tilde{\ell}_Z\left(\bm{\theta};\{h( v_{j,i},t_{j,i};\bm{\theta}^\star )\}_{j=1,i=0}^{M,N}\right).
$
\State \textbf{repeat} steps (2 to 6) \textbf{until} the numerical approximation to the solution of the fixed-point problem (\ref{FP}) is found.
\end{algorithmic}
\end{algorithm}

\begin{figure}[H]
\centering
\captionsetup{skip=4pt}
\includegraphics[width=0.485\textwidth]{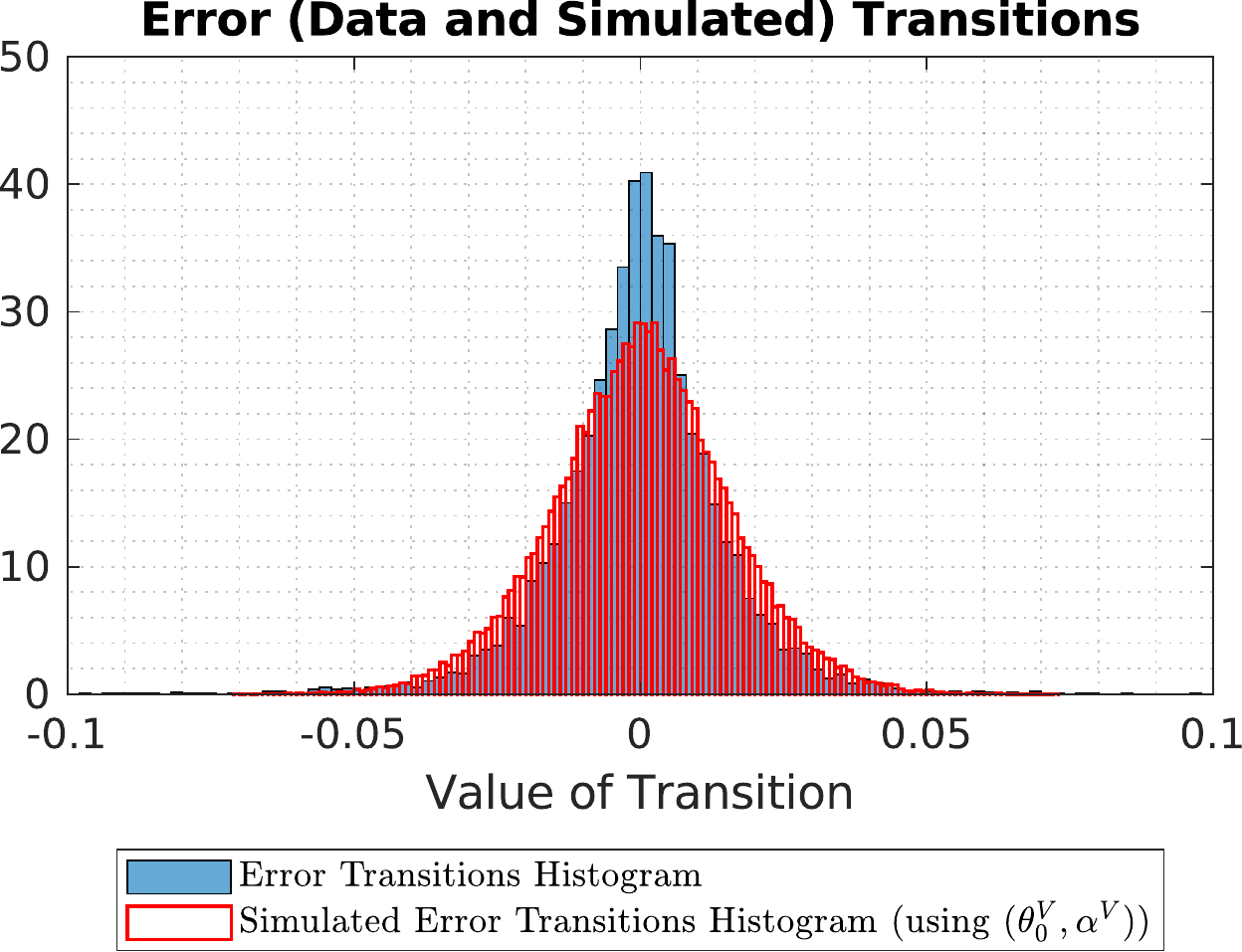}\quad
\includegraphics[width=0.485\textwidth]{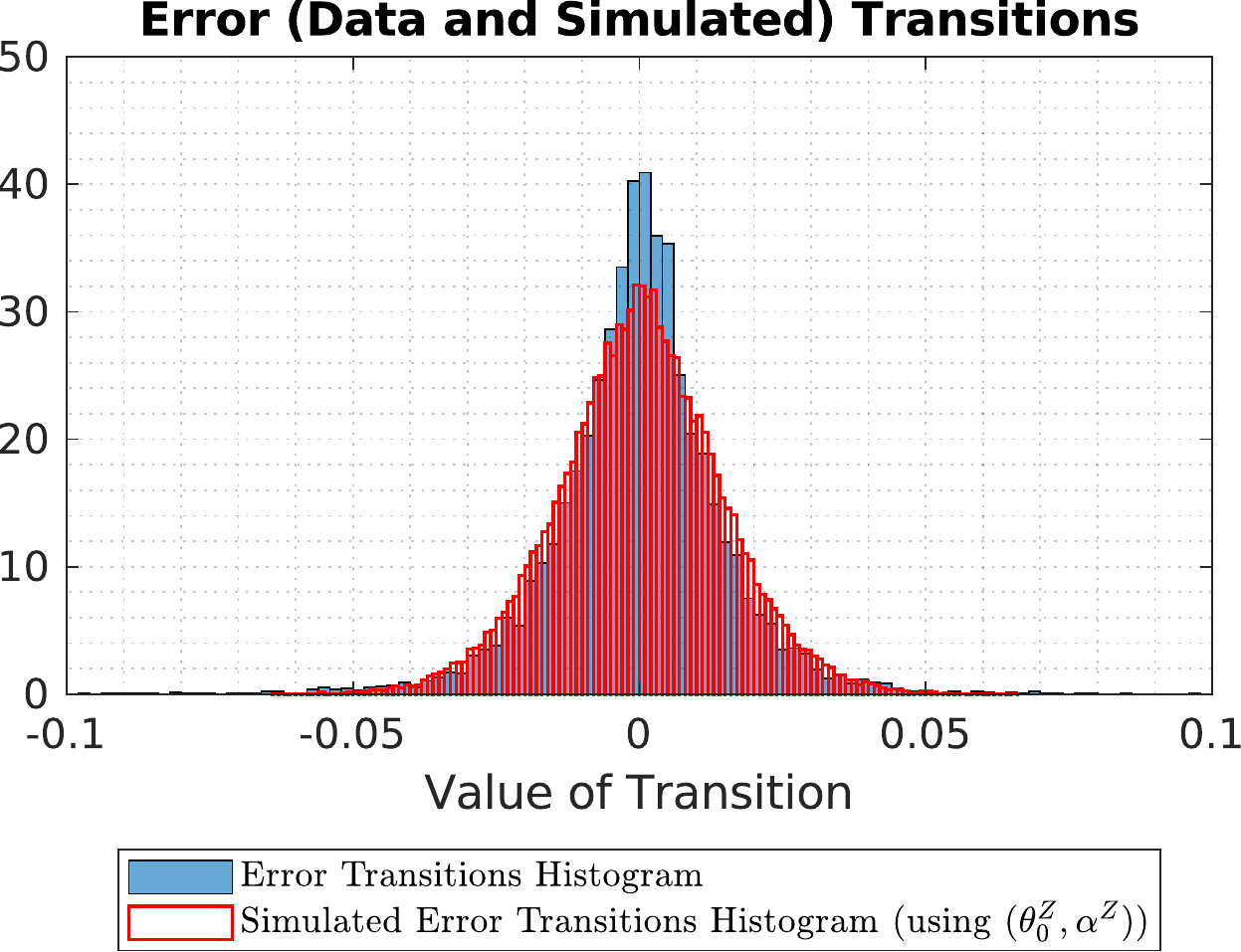}
\caption{Density histograms for error transitions. Using provider A, we overlap the real transitions from the test set with the simulated ones from the $V-$space SDE. On the left, simulations use $(\theta_0^V,\alpha^V)$. On the right, simulations use $(\theta_0^Z,\alpha^Z)$.}
\label{fig:hists}
\end{figure}

To verify and compare these two vector of parameters, (i.e., $(\theta_0^V,\alpha^V)$ and $(\theta_0^Z,\alpha^Z)$), we simulate error paths in the $V-$space. We simulate five error paths for each day in the test set and construct histograms with the transitions. The histograms can be seen in Figure (\ref{fig:hists}). We observe a slightly better approximation using $(\theta^Z_0,\alpha^Z)$.

\vspace{-0.3cm}
\subsection{Model comparison and assessment of the forecast providers} \label{Model_Comp}

We compare two candidate models to find the best-fit that maximizes the retained information, the Model 1, introduced in (\cite{elkate}, p.383), and our proposed model  (\ref{ourmodel}), from hereafter called Model 2.
\begin{itemize}
  \item Model 1: This model does not feature derivative tracking: 
\begin{equation}
  \left\{
  \begin{array}{@{}rl@{}}
    dX_t \!\!\!&=  -\theta_0 (X_t - p_t) d t + \sqrt{2 \alpha \theta_0 X_t (1 - X_t)} dW_t, \:\: t \in [0,T]  \\
     X_0  \!\!\!&=  x_0 \in [0,1], 
  \end{array}  \right. \label{Model1}  
\end{equation}  
 with $\theta_0 > 0, \, \alpha > 0$.
\end{itemize}

\begin{itemize}
  \item Model 2: This model features derivative tracking and time-varying mean-reversion parameter:  
\begin{equation}
  \left\{
  \begin{array}{@{}rl@{}}
    dX_t \!\!\!&= \big(\dot{p}_t  - \theta_t (X_t - p_t) \big) d t +\sqrt{2 \alpha \theta_0 X_t (1 - X_t)} dW_t, \:\: t \in [0,T]\\
     X_0  \!\!\!&=  x_0 \in [0,1],
  \end{array}\right. \label{Model2}
\end{equation}  
 with $\theta_0 > 0$, $\alpha > 0$ and $\theta_t$ satisfying condition \eqref{Assumption:3} .
\end{itemize}
To show the better performance of Model 2, we have computed the Akaike information criterion (AIC) and the Bayesian information criterion (BIC) for the two considered models, and any combination of the three different forecast providers with  three approximate likelihood methods, the one based on the Beta surrogate density in the $V$-space (Subsection \ref{moments_ODEs}), the one based on the Gaussian surrogate density in the $Z$-space (Subsection \ref{moments_ODEs_Z}), and the Shoji-Ozaki local linearization method (\cite{shoz}). 
Table (\ref{tab:model_comparison}) summarizes these results, also reporting the estimate of the variability diffusion coefficient $\alpha \theta_0$. 
It is worth observing that the best fitting is achieved with Model 2 and adopting Beta distributions as proxies of the transition densities.
\begingroup
\renewcommand{\arraystretch}{0.8}
\begin{table}[h]
\caption{Model comparison based on Akaike and Bayesian information criteria.}
\label{tab:model_comparison}
\centering
\begin{tabular}{ c @{\hspace{1\tabcolsep}} c @{\hspace{0.8\tabcolsep}} c @{\hspace{0.8\tabcolsep}} c  @{\hspace{0.8\tabcolsep}} cc}
\hline\noalign{\smallskip}
Model & \begin{tabular}{@{}c@{}}Forecast \\ Provider\end{tabular} & Method & \begin{tabular}{@{}c@{}}Product \\ $\theta_0\alpha$ \end{tabular}  & AIC & BIC \\ 
\hline\noalign{\smallskip}
Model 1 & Provider A & Gaussian Proxy & 0.105 & -58226 & -58211 \\
 &  & Shoji-Ozaki & 0.104 & -58226 & -58211 \\
 &  & Beta Proxy & 0.104 & -58286 & -58271 \\
 & Provider B & Gaussian Proxy & 0.105 & -58226   & -58211 \\
 &  & Shoji-Ozaki & 0.104 & -58226 & -58211 \\
 &  & Beta Proxy & 0.104 & -58288 & -58273 \\
 & Provider C & Gaussian Proxy & 0.105 & -58226 & -58211 \\
 &  & Shoji-Ozaki & 0.104 & -58226 & -58211 \\
 &  & Beta Proxy & 0.104 & -58286 & -58271 \\
Model 2 & Provider A & Beta Proxy & 0.097 & -73700   & -73685 \\ 
 & Provider B & Beta Proxy & 0.098 &  -73502 & -73487 \\ 
 & Provider C & Beta Proxy & 0.108 & -72518 & -72503 \\ 
\noalign{\smallskip}\hline
\end{tabular}
\end{table}
\endgroup

The optimal estimates of the parameters of Model 2, for the three forecast providers, when using Beta surrogates for the transition density are presented next:
\begingroup
\renewcommand{\arraystretch}{0.8}
\begin{table}[H]
\caption{Optimal parameters for the three different forecast providers using Model 2 with Beta proxies.}
\label{tab:forcast_comparison}
\centering
\begin{tabular}{ccc}
\hline\noalign{\smallskip}
Forecast Provider & Parameters $(\theta_0, \alpha)$ & Product $\theta_0\alpha$ \\ 
\hline\noalign{\smallskip}
Provider A  & $(1.93,0.050)$  &  0.097 \\
Provider B  & $(1.42,0.069) $  &  0.098 \\ 
Provider C  & $(1.38,0.078) $  &  0.108 \\ 
\noalign{\smallskip}\hline
\end{tabular}
\end{table}
\endgroup

\subsection{Calibration of Model 2 with additional parameter $\delta$}

\begin{figure}[H]
\centering
\includegraphics[width=0.485\textwidth]{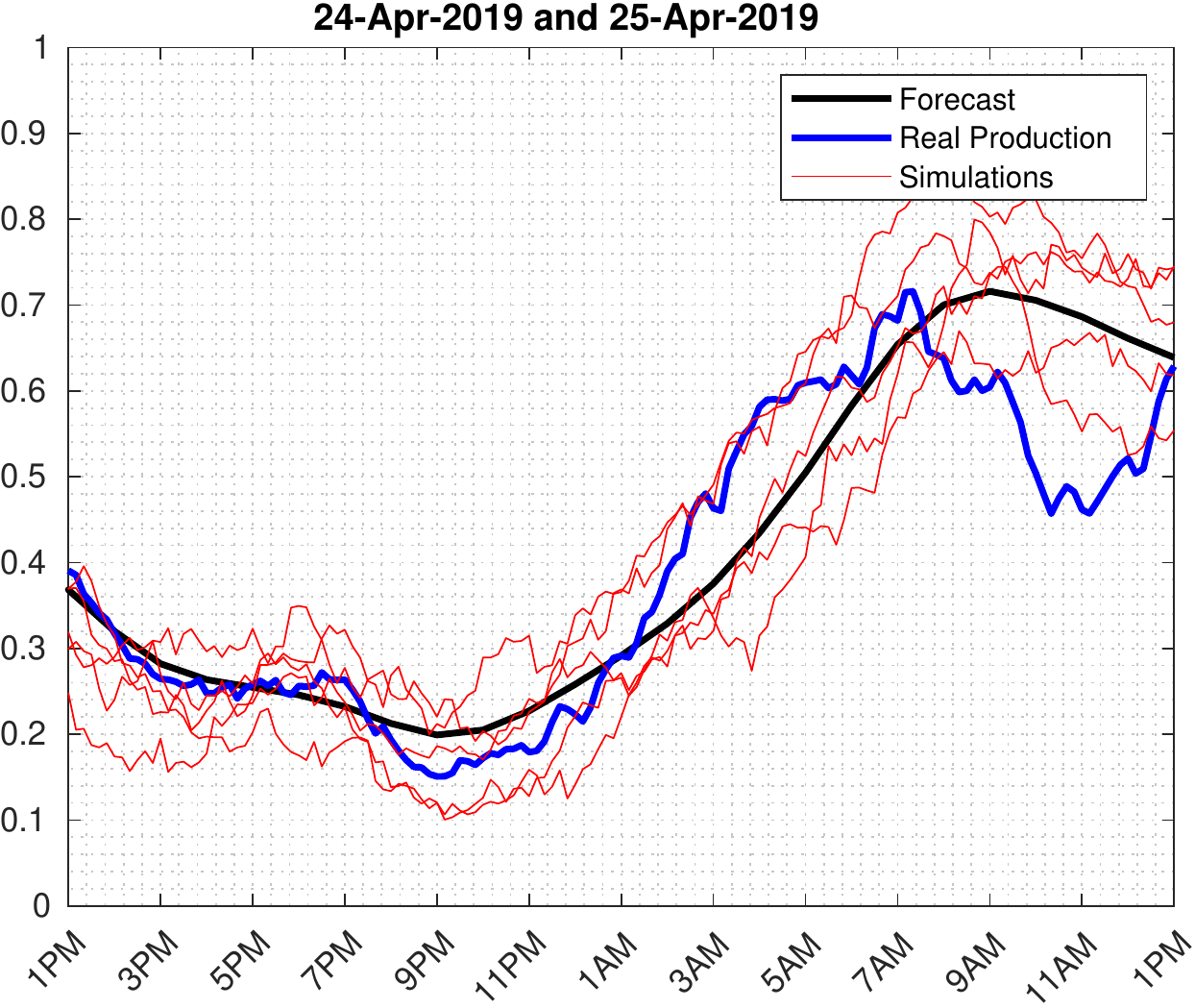}\quad
\includegraphics[width=0.485\textwidth]{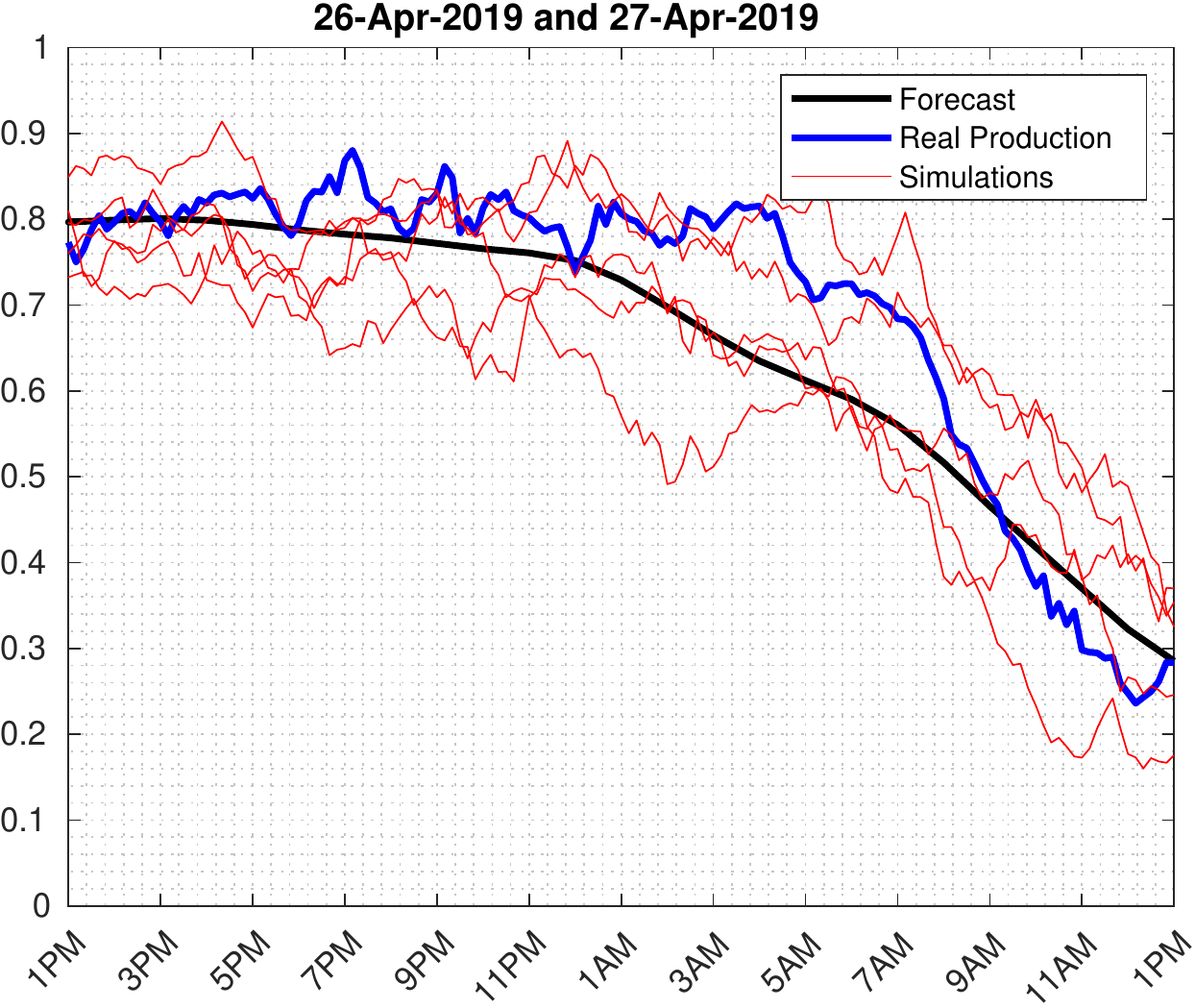}\\
\caption{Two arbitrary days with five simulated wind power production paths each.}
\label{fig:simulation_paths}
\end{figure}

After calibrating Model 2 on the training set using the complete likelihood (\ref{eq:complete_LH}), we can generate simulations of the wind power production for the time horizon of interest. Figure (\ref{fig:simulation_paths}) shows five simulated paths of wind power production for each day of interest.

\begin{figure}[H]
\centering
\captionsetup{skip=4pt}
\includegraphics[width=0.485\textwidth]{plots/bands/1.pdf}\quad
\includegraphics[width=0.485\textwidth]{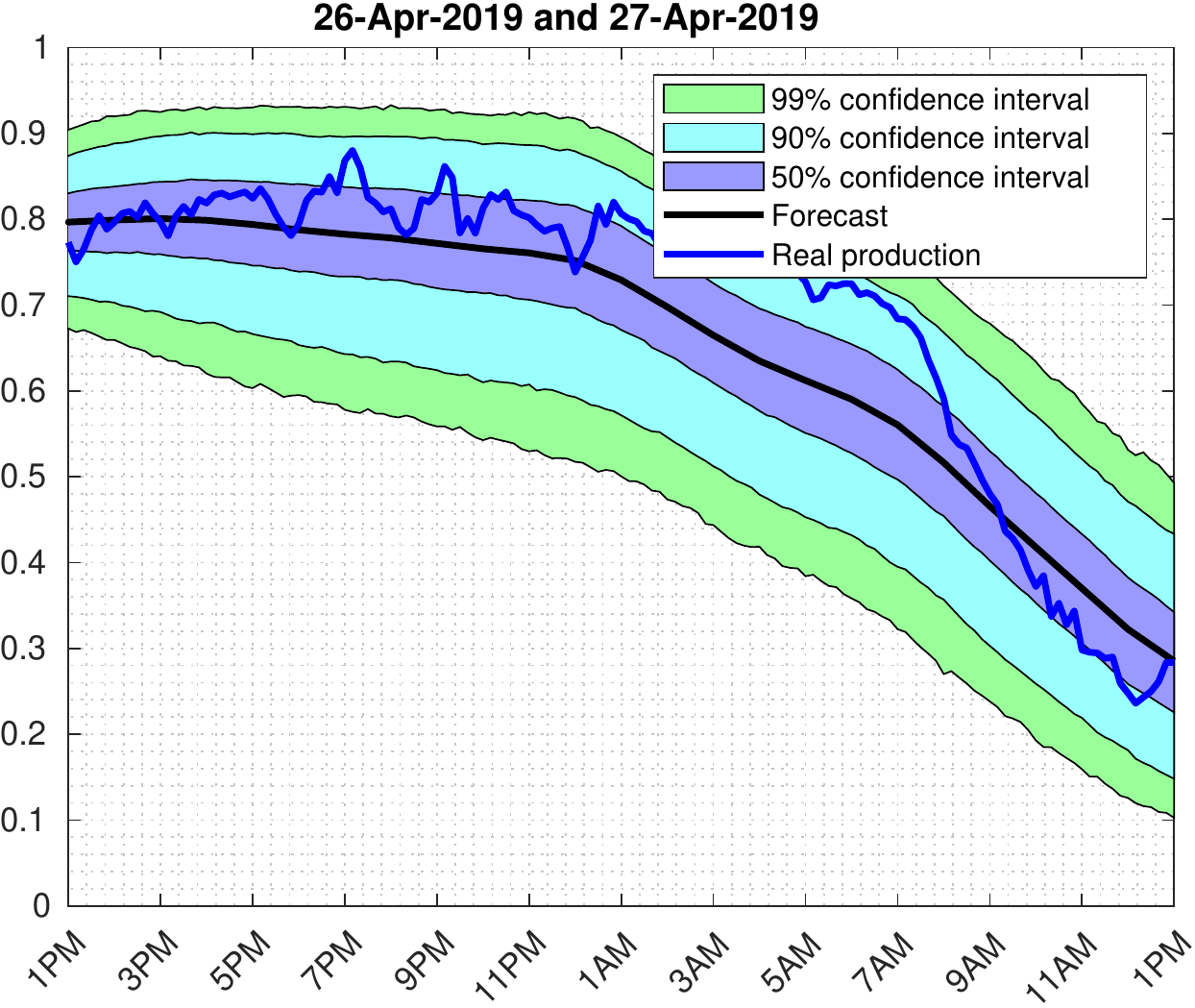}\\
\caption{Empirical pointwise confidence bands for the wind power production using the approximate MLEs for Model 2 $(\theta_0, \alpha ,\delta)=(2.22,0.044,0.054)$. Blue line: real production.}
\label{fig:confidence_bands}
\end{figure}

Once derived optimal estimates of the parameters of the complete likelihood for Model 2, we obtain empirical pointwise confidence bands for wind power production. Figure (\ref{fig:confidence_bands}) shows the empirical pointwise confidence bands for wind power production for each day of interest, assuming Model 2 specification, a given forecaster, and 5000 simulations per day.

\subsubsection{Value of $\delta$}
As a final verification, we study the behavior of $\delta$ as a function of the vector $\bm{\theta}$. Given a parameter vector, we calculate an initial guess for $\delta$ solving problem (\ref{eq:likelihood_delta}). Even when it is a guess, it helps us understand the meaning of this additional parameter qualitatively.

\begin{figure}[H]
\centering
\captionsetup{skip=2pt}
\includegraphics[width=0.70\textwidth]{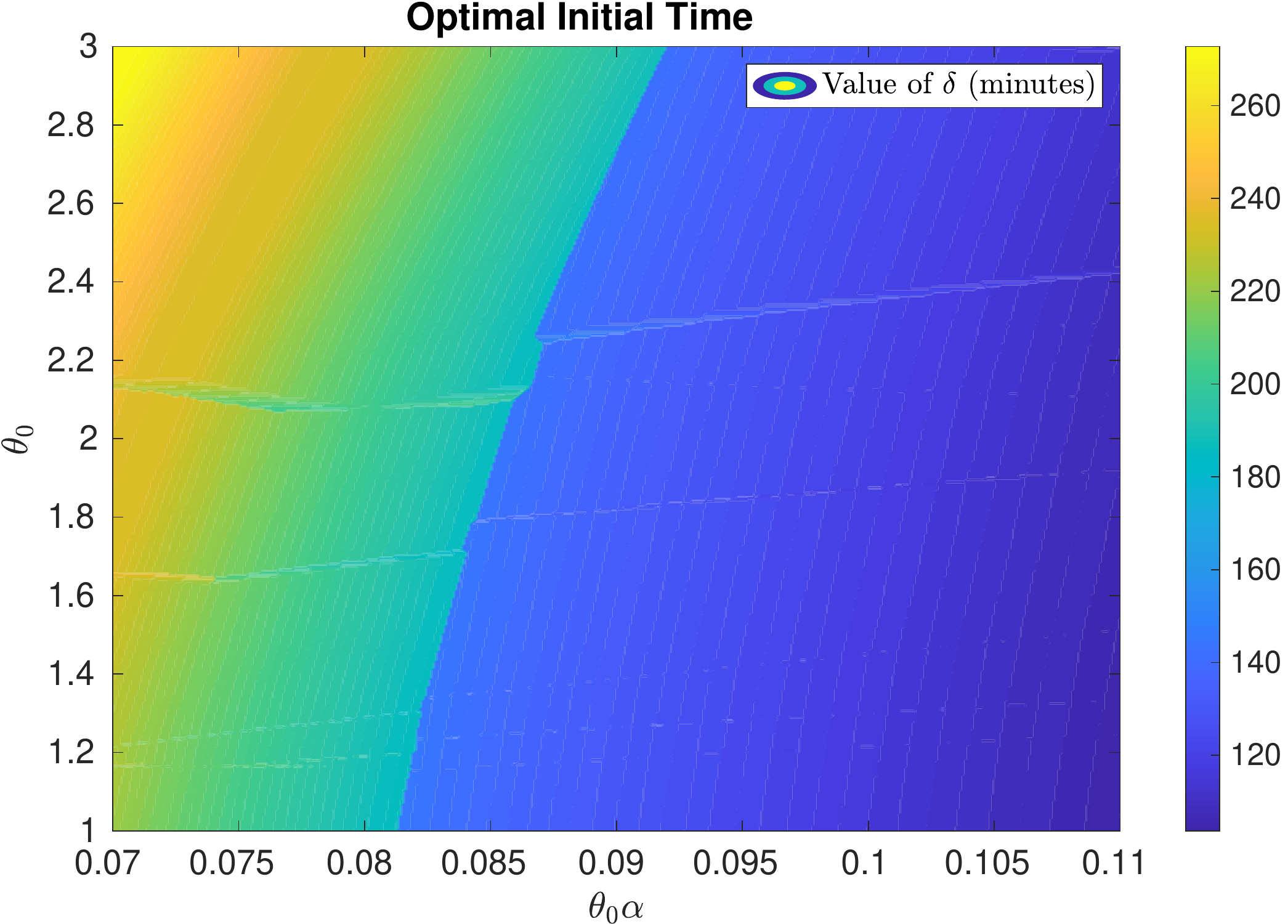}
\caption{Initial value for $\delta$ as a function of the elements of the parameter vector $\bm{\theta}$.}
\label{fig:delta}
\end{figure}

We choose as domain the most significant values of $\theta_0$ and $\theta_0\alpha$, regarding the previous numerical results. In Figure (\ref{fig:delta}) we can observe that:
\begin{itemize}
\item The initial time $\delta$ decreases as $\theta_0\alpha$ increases. This is a consequence of the increment in the diffusion as $\theta_0\alpha$ increases. As there is more diffusion, less time is needed for the initial transition density to cover the initial error observations.
\item The initial time $\delta$ increases as $\theta_0$ increases. As we increment $\theta_0$, the mean reversion becomes larger and reduces the variance for the initial transition density. Then, more time is needed for the initial transition density to cover the initial error observations.\end{itemize}



\section{Conclusions} \label{Section_6}

We have developed a methodology for assessing forecast uncertainty, which is agnostic of the forecasting technology and applicable to real-world problems where historical observations and their forecasts are available.

To this purpose, we built a data-driven stochastic differential equation model for the normalized forecast error, with time-varying mean-reversion parameter in the linear drift coefficient, and state-dependent and time non-homogenous diffusion coefficient. We also used the Lamperti transform with unknown parameters to provide a version of the proposed model with a unit diffusion coefficient, increasing its stability properties.

We used approximate likelihood-based methods for the models' calibration, both in the original forecast and the Lamperti space, relying on moment-matching techniques that require solving systems of ordinary differential equations. 
For the Lamperti space, we derived optimal estimates of the unknown parameters using a novel fixed-point optimization procedure. 

The likelihood approach allowed for the extending of the SDE models in a very effective way, incorporating an early transition with an additional parameter that accounts for the forecast's uncertainty at the beginning of each future period. As a result, we obtained a robust procedure for synthetic data generation that, using the available forecast input, embraces future sample paths through empirical pointwise bands with prescribed confidence. 

On the basis of historical data of wind power production and forecast from different sources, our method came up with an objective tool for forecast assessment and comparison by performing the model selection stage.
The application of the modeling procedure, inference through numerical optimization, and model selection through information criteria, to the wind power production dataset in Uruguay between April and December 2019, with three different pro\-vid\-ers, shows the excellent performance of our proposed model, which preserves the asymmetry of wind power forecast errors and their correlation structure.
 
We conclude that our SDE model, featuring a time-de\-riv\-a\-tive tracking of the forecast, a time-dependent mean-reversion parameter, and a state-dependent diffusion term that suitably adjusts to the problem under study, contributes efficiently toward the management of several types of data, such as renewable energies. Future work will address solar power forecast pathwise uncertainty quantification, where is required the estimation of the daily maximum solar power production to get a realistic time-varying upper bound for the path variability. This methodology paves the way for stochastic optimal control methods enabling principled decision making under uncertainty in the presence of complex data matrices.
 

\section{Appendix}
\label{sec:app}

For a time horizon $T>0$, a parameter $\alpha > 0$, and $(\theta_t)_{t\in[0,T]}$ a positive deterministic  function,  let us consider the model  given by
\begin{equation}
  \left\{
  \begin{array}{@{}rl@{}}
    dX_t \!\!\!&= \big(\dot p_t - \theta_t (X_t-p_t)  \big) d t  +\sqrt{2 \alpha \theta_0 X_t (1-X_t)} dW_t, \:\: t \in [0,T]\\
     X_0  \!\!\!&=  x_0 \in [0,1],
  \end{array}\right. \label{eq:model}
\end{equation} 
where $(p_t)_{t\in[0,T]}$ denotes the prediction function that satisfies $0\le p_t\le 1$ for all $t\in[0,T]$. This prediction function is assumed to be a smooth function of the time so that  \vspace{-0.2cm}
$$\sup_{t\in[0,T]}\bigl( |p_s| + |\dot p_s|\big) <+\infty .$$
The following proofs are based on standard arguments for stochastic processes that can be found e.g. in \cite{Alf} and \cite{KarShr} that we adapted to the setting of our model \eqref{eq:model}.
\begin{thm}\label{thm:exun}
Assume that    
\begin{equation}\label{Assumption:1}
\forall  t\in[0,T],\;\; 0\le \dot p_t +\theta_tp_t\le \theta_t, \;\;\mbox{ and }\;\;
\sup_{t\in[0,T]}|\theta_t|<+\infty\tag{A}. 
\end{equation}
Then, there is a unique strong solution to \eqref{eq:model} s.t.  for all $t\in[0,T]$, $X_t\in[0,1]$ a.s.
\end{thm}
\begin{pf}
Let us first consider the following SDE for $t\in[0,T]$
\begin{align}\label{eq:eds1}
X_t &=x_0+  \int_0^t\big(\dot p_s - \theta_s(X_s-p_s)  \big) d s  \nonumber \\ 
& + \int_0^t\sqrt{2\alpha \theta_0 |X_s(1-X_s)|} dW_s, \quad 0\leq x_0\leq1.
\end{align}
According to Proposition 2.13, p.291 of \cite{KarShr}, under assumption  \eqref{Assumption:1} there is a unique strong solution $X_t$ to \eqref{eq:eds1}. Moreover, as the diffusion coefficient is of linear growth, we have  for all $p>0$ 
\begin{equation}\label{prop:fm}
\mathbb E\left[ \sup_{t\in[0,T]}|X_t|^p\right]<\infty.
\end{equation}
Then, it remains to show that for all $t\in[0,T]$, $X_t\in[0,1]$ a.s. For this aim, we need to use the so-called Yamada function $\psi_n$ that is a $\mathcal C^2$ function that satisfies a bunch of useful properties:
\begin{align*}
&|\psi_n(x)|\underset{n\rightarrow+\infty}{\rightarrow}|x|, \;\; x{\psi'}_n(x)\underset{n\rightarrow+\infty}{\rightarrow}|x|, \\  
&|\psi_n(x)|\wedge |x{\psi'}_n(x)| \le |x|, \;\; {\psi'}_n(x) \le 1, \\ & \mbox{and}\;\; {\psi''}_n(x)=g_n(|x|)\ge 0\;\; \mbox{with} \;\; g_n(x)x\le \frac 2n\;\;  \mbox{for all}\;\; x\in \mathbb R .
\end{align*}
See the proof of Proposition 2.13, p. 291 of \cite{KarShr} for the construction of such function.
Applying Itô's formula we get
\begin{align*}
\psi_n(X_t)&=\psi_n(x_0) +\int_0^t {\psi'}_n(X_s)(\dot p_s + \theta_s p_s - \theta_sX_s  \big) ds \\
&+ \int_0^t{\psi'}_n(X_s)\sqrt{2\alpha \theta_0 |X_s(1-X_s)|} dW_s \\ 
&+ \alpha \theta_0 \int_0^t  g_n(|X_s|) |X_s(1-X_s)|  ds.
\end{align*}
Now, thanks to  \eqref{Assumption:1}, \eqref{prop:fm}, and to the above properties of $\psi_n$ and $g_n$, we get
\begin{align*}
\mathbb E\left[\psi_n(X_t)\right] &\le \psi_n(x_0) +\int_0^t \left(\dot p_s + \theta_sp_s -  \theta_s \mathbb E[{\psi'}_n(X_s)X_s] \right) ds 
\\ 
&+ \frac{2\alpha  \theta_0}{n}\int_0^t \mathbb E \left[|1-X_s|\right] ds.
\end{align*}
Therefore, letting $n$ tends to infinity, we use Lebesgue's theorem to get
$$
\mathbb E\left[|X_t|\right]\le x_0 +\int_0^t \left(\dot p_s + \theta_sp_s -  \theta_s \mathbb E\left[|X_s|\right] \right) ds.
$$
Besides, taking the expectation of \eqref{eq:eds1}, we get
$$
\mathbb E \left[X_t\right]=x_0+  \int_0^t\big(\dot p_s +\theta_sp_s - \theta_s \mathbb E\left[X_s\right]  \big) ds,
$$
and thus we have 
$$
\mathbb E\left[|X_t| -X_t \right]\le \int_0^t \theta_s \mathbb E\left[ X_s - |X_s| \right] ds.
$$
Then, Gronwall's lemma gives us $\mathbb E\left[|X_t|\right]=\mathbb E \left[X_t\right]$ and thus for any $t\in[0,T]$ $X_t\ge0$ a.s. The same arguments work to prove that  for any $t\in[0,T]$ $Y_t:=1-X_t\ge0$  a.s.  since the process $(Y_t)_{t\in[0,T]}$ is solution to 
$$
dY_t= \big( \theta_t(1-p_t) -\dot p_t - \theta_tY_t  \big) dt  -\sqrt{2\alpha \theta_0Y_t(1-Y_t)} dW_t \,.
$$
Then similarly, we need to assume that $\dot p_t +\theta_tp_t\ge 0$. This completes the proof.
\end{pf}

\begin{thm} \label{thm:mod2}
Assume that assumptions of Theorem \ref{thm:exun} hold with $x_0\in]0,1[$.
Let $\tau_0:=\inf \{t\in[0,T],\; X_t=0\}$ and  $\tau_1:=\inf \{t\in[0,T],\; X_t=1\}$ with the convention that $\inf\emptyset=+\infty$. Assume in addition that for all $t\in[0,T]$,  $p_t\in]0,1[$ and that
\begin{equation}\label{Assumption:3}
\theta_t\geq \max\left(\frac{\alpha\theta_0+\dot p_t}{1-p_t},\frac{\alpha\theta_0-\dot p_t}{p_t}\right)\tag{B}. 
\end{equation}
 Then, $\tau_0=\tau_1=+\infty$ a.s.
\end{thm}

\begin{pf}
For $t\in[0,\tau_0[$, we have 
$$
\frac{dX_t}{X_t}= \left(\frac{\dot p_t +\theta_t p_t}{X_t} - \theta_t\right) dt  +\sqrt{\frac{2\alpha \theta_0 (1-X_t)}{X_t}} dW_t 
$$  
so that
$$
X_t=x_0\exp\left(\int_0^t \frac{\dot p_s +\theta_sp_s- \theta_0 \alpha}{X_s} ds+\alpha \theta_0  t-  \int_0^t\theta_s ds + M_t\right),
$$
where $M_t=\int_0^t\sqrt{\frac{2\alpha \theta_0 (1-X_s)}{X_s}} dW_s$ is a continuous martingale. Then, as for all $t\in[0,T]$ we have $\dot p_t +\theta_tp_t- \theta_0 \alpha\ge0$, we deduce that
$$
X_t\ge x_0\exp\left(\alpha\theta_0 t-  \int_0^t\theta_s ds + M_t\right).
$$
By way of contradiction let us assume that  $\{\tau_0<\infty\}$, then letting $t\to \tau_0$ we deduce that $$\lim_{t\to \infty} {\mathbf 1}_{\{\tau_0<\infty\}}M_{t\wedge \tau_0}= - {\mathbf 1}_{\{\tau_0<\infty\}}\infty \: \textrm{ a.s.} $$ This leads to a contradiction since we know that continuous martingales likewise the Brownian motion cannot converge almost surely to $+\infty$ or $-\infty$. It follows that $\tau_0=\infty$ almost surely. Next, recalling that  the process $(Y_t)_{t\geq 0}$  given by $Y_t=1-X_t$ is solution to
$$
dY_t= \big( \theta_t(1-p_t) -\dot p_t - \theta_t Y_t  \big) dt  -\sqrt{2 \alpha \theta_0 Y_t(1-Y_t)} dW_t,
$$
we deduce using similar arguments as above 
$\tau_1=\infty$ a.s. provided that $\theta_t(1-p_t) -\dot p_t -\alpha \theta_0 \ge 0$.
\end{pf} 

\begin{rmk} 
As the diffusion coefficient of $X_t$  given by  $x \mapsto \sqrt{2 \alpha \theta_0 x(1-x)}$  is strictly positive for all $x \in  ]0,1[$, the condition \eqref{Assumption:3}  ensures that the transformation between $Z_t$ and $X_t$ is bijective, so that we deduce the properties of existence and uniqueness of $Z_t$ from those of $X_t$. The application of  It\^{o}'s formula in Section \ref{Section_4} is subjected to the condition \eqref{Assumption:3} that avoids the process $X_t$ hits the boundaries of the interval $ ]0,1[$, otherwise the Lamperti transform is not applicable.
\end{rmk}

\section*{Acknowledgements}
This research was partially supported by the KAUST Office of Sponsored Research (OSR) under Award number URF/1/2584 -- 01 -- 01 in the KAUST Competitive Research Grants Program Round 8, the Alexander von Humboldt Foundation, the chair Risques Financiers, Fondation du Risque, and the Laboratory of Excellence MME-DII Grant no. ANR11-LBX--0023--01 (\url{http://labex-mme-dii.u-cergy.fr/}). We thank UTE (\url{https://portal.ute.com.uy/}) for providing the data used in this research.

\bibliography{mybibfile}

\begin{thebibliography}{23}
\expandafter\ifx\csname natexlab\endcsname\relax\def\natexlab#1{#1}\fi
\providecommand{\url}[1]{\texttt{#1}}
\providecommand{\href}[2]{#2}
\providecommand{\path}[1]{#1}
\providecommand{\DOIprefix}{doi:}
\providecommand{\ArXivprefix}{arXiv:}
\providecommand{\URLprefix}{URL: }
\providecommand{\Pubmedprefix}{pmid:}
\providecommand{\doi}[1]{\href{http://dx.doi.org/#1}{\path{#1}}}
\providecommand{\Pubmed}[1]{\href{pmid:#1}{\path{#1}}}
\providecommand{\bibinfo}[2]{#2}
\ifx\xfnm\relax \def\xfnm[#1]{\unskip,\space#1}\fi
\bibitem[{A{\"i}t-Sahalia(2002)}]{ait}
\bibinfo{author}{A{\"i}t-Sahalia, Y.}, \bibinfo{year}{2002}.
\newblock \bibinfo{title}{Maximum likelihood estimation of discretely sampled
  diffusions: a closed-form approximation approach}.
\newblock \bibinfo{journal}{Econometrica} \bibinfo{volume}{70},
  \bibinfo{pages}{223--262}.
\newblock \DOIprefix\doi{10.1111/1468-0262.00274}.
\bibitem[{Alfonsi(2015)}]{Alf}
\bibinfo{author}{Alfonsi, A.}, \bibinfo{year}{2015}.
\newblock \bibinfo{title}{Affine {D}iffusions and {R}elated {P}rocesses:
  {S}imulation, {T}heory and {A}pplications}. volume~\bibinfo{volume}{6} of
  \textit{\bibinfo{series}{Bocconi \& Springer Series}}.
\newblock \bibinfo{publisher}{Springer}.
\newblock \DOIprefix\doi{10.1007/978-3-319-05221-2}.
\bibitem[{Badosa et~al.(2018)Badosa, Gobet, Grangereau and Kim}]{bggk}
\bibinfo{author}{Badosa, J.}, \bibinfo{author}{Gobet, E.},
  \bibinfo{author}{Grangereau, M.}, \bibinfo{author}{Kim, D.},
  \bibinfo{year}{2018}.
\newblock \bibinfo{title}{Day-ahead probabilistic forecast of solar irradiance:
  A stochastic differential equation approach}, in: \bibinfo{editor}{Drobinski,
  P.}, \bibinfo{editor}{Mougeot, M.}, \bibinfo{editor}{Picard, D.},
  \bibinfo{editor}{Plougonven, R.}, \bibinfo{editor}{Tankov, P.} (Eds.),
  \bibinfo{booktitle}{Renewable Energy: Forecasting and Risk Management},
  \bibinfo{publisher}{Springer International Publishing},
  \bibinfo{address}{Cham}. pp. \bibinfo{pages}{73--93}.
\newblock \DOIprefix\doi{10.1007/978-3-319-99052-1_4}.
\bibitem[{D'Onofrio et~al.(2018)D'Onofrio, Tamborrino and Lansky}]{dotala}
\bibinfo{author}{D'Onofrio, G.}, \bibinfo{author}{Tamborrino, M.},
  \bibinfo{author}{Lansky, P.}, \bibinfo{year}{2018}.
\newblock \bibinfo{title}{The {J}acobi diffusion process as a neuronal model}.
\newblock \bibinfo{journal}{Chaos} \bibinfo{volume}{28}.
\newblock \DOIprefix\doi{10.1063/1.5051494}.
\bibitem[{Egorov et~al.(2003)Egorov, Li and Xu}]{eglix}
\bibinfo{author}{Egorov, A.V.}, \bibinfo{author}{Li, H.}, \bibinfo{author}{Xu,
  Y.}, \bibinfo{year}{2003}.
\newblock \bibinfo{title}{Maximum likelihood estimation of time-inhomogeneous
  diffusions}.
\newblock \bibinfo{journal}{Journal of Econometrics} \bibinfo{volume}{114},
  \bibinfo{pages}{107--139}.
\newblock \DOIprefix\doi{10.1016/S0304-4076(02)00221-X}.
\bibitem[{Elkantassi et~al.(2017)Elkantassi, Kalligiannaki and
  Tempone}]{elkate}
\bibinfo{author}{Elkantassi, S.}, \bibinfo{author}{Kalligiannaki, E.},
  \bibinfo{author}{Tempone, R.}, \bibinfo{year}{2017}.
\newblock \bibinfo{title}{Inference and {S}ensitivity in {S}tochastic {W}ind
  {P}ower {F}orecast {M}odels}, in: \bibinfo{editor}{Papadrakakis, M.},
  \bibinfo{editor}{Papadopoulos, V.}, \bibinfo{editor}{Stefanou, G.} (Eds.),
  \bibinfo{booktitle}{2nd {ECCOMAS} {T}hematic {C}onference on {U}ncertainty
  {Q}uantification in {C}omputational {S}ciences and {E}ngineering},
  \bibinfo{publisher}{Eccomas Proceedia UNCECOMP 2017}. pp.
  \bibinfo{pages}{381--393}.
\newblock \DOIprefix\doi{10.7712/120217.5377.16899}.
\bibitem[{Forman and Sorensen(2008)}]{foso}
\bibinfo{author}{Forman, J.L.}, \bibinfo{author}{Sorensen, M.},
  \bibinfo{year}{2008}.
\newblock \bibinfo{title}{The {P}earson {D}iffusions: {A} {C}lass of
  {S}tatistically {T}ractable {D}iffusion {P}rocesses}.
\newblock \bibinfo{journal}{Scandinavian Journal of Statistics}
  \bibinfo{volume}{35}, \bibinfo{pages}{438--465}.
\newblock \DOIprefix\doi{10.1111/j.1467-9469.2007.00592.x}.
\bibitem[{Iacus(2008)}]{iacus1}
\bibinfo{author}{Iacus, S.M.}, \bibinfo{year}{2008}.
\newblock \bibinfo{title}{Simulation and {I}nference for {S}tochastic
  {D}ifferential {E}quations: {W}ith {R} {E}xamples}.
\newblock Springer Series in Statistics, \bibinfo{publisher}{Springer},
  \bibinfo{address}{New York}.
\newblock \DOIprefix\doi{10.1007/978-0-387-75839-8}.
\bibitem[{IRENA(2018)}]{irena2}
\bibinfo{author}{IRENA}, \bibinfo{year}{2018}.
\newblock \bibinfo{title}{Uruguay {P}ower {S}ystem {F}lexibility assessment:
  IRENA {F}lex{T}ool {C}ase {S}tudy}.
\newblock \bibinfo{address}{Abu Dhabi}.
\bibitem[{IRENA(2019)}]{irena}
\bibinfo{author}{IRENA}, \bibinfo{year}{2019}.
\newblock \bibinfo{title}{Innovation landscape for a renewable-powered future:
  Solutions to integrate variable renewables}.
\newblock \bibinfo{address}{Abu Dhabi}.
\bibitem[{Iversen et~al.(2014)Iversen, Morales, M{\o}ller and Madsen}]{immm}
\bibinfo{author}{Iversen, E.B.}, \bibinfo{author}{Morales, J.M.},
  \bibinfo{author}{M{\o}ller, J.K.}, \bibinfo{author}{Madsen, H.},
  \bibinfo{year}{2014}.
\newblock \bibinfo{title}{Probabilistic forecasts of solar irradiance using
  stochastic differential equations}.
\newblock \bibinfo{journal}{Environmetrics} \bibinfo{volume}{25},
  \bibinfo{pages}{152--164}.
\newblock \DOIprefix\doi{10.1002/10.1002/env.2267}.
\bibitem[{Karatzas and Shreve(1998)}]{KarShr}
\bibinfo{author}{Karatzas, I.}, \bibinfo{author}{Shreve, S.E.},
  \bibinfo{year}{1998}.
\newblock \bibinfo{title}{Brownian motion}, in: \bibinfo{booktitle}{Brownian
  Motion and Stochastic Calculus}. \bibinfo{publisher}{Springer New York},
  \bibinfo{address}{New York, NY}, pp. \bibinfo{pages}{47--127}.
\newblock \DOIprefix\doi{10.1007/978-1-4612-0949-2_2}.
\bibitem[{Lamperti(1964)}]{lamp}
\bibinfo{author}{Lamperti, J.}, \bibinfo{year}{1964}.
\newblock \bibinfo{title}{A simple construction of certain diffusion
  processes}.
\newblock \bibinfo{journal}{J. Math. Kyoto Univ.} \bibinfo{volume}{4},
  \bibinfo{pages}{161--170}.
\newblock \DOIprefix\doi{10.1215/kjm/1250524711}.
\bibitem[{Leonenko and Phillips(2012)}]{leph}
\bibinfo{author}{Leonenko, G.}, \bibinfo{author}{Phillips, T.},
  \bibinfo{year}{2012}.
\newblock \bibinfo{title}{High-order approximation of {P}earson diffusions
  processes}.
\newblock \bibinfo{journal}{Journal of Computational and Applied Mathematics}
  \bibinfo{volume}{236}, \bibinfo{pages}{2853--2868}.
\newblock \DOIprefix\doi{10.1016/j.cam.2012.01.022}.
\bibitem[{M{\o}ller and Madsen(2010)}]{moma}
\bibinfo{author}{M{\o}ller, J.K.}, \bibinfo{author}{Madsen, H.},
  \bibinfo{year}{2010}.
\newblock \bibinfo{title}{From {S}tate {D}ependent {D}iffusion to {C}onstant
  {D}iffusion in {S}tochastic {D}ifferential {E}quations by the {L}amperti
  {T}ransform}.
\newblock \bibinfo{type}{Technical Report} \bibinfo{number}{IMM-Technical
  Report-2010-16}. Technical University of Denmark, DTU Informatics, Building
  321. \bibinfo{address}{Kgs. Lyngby, Denmark}.
\bibitem[{M{\o}ller et~al.(2016)M{\o}ller, Zugno and Madsen}]{mozuma}
\bibinfo{author}{M{\o}ller, J.K.}, \bibinfo{author}{Zugno, M.},
  \bibinfo{author}{Madsen, H.}, \bibinfo{year}{2016}.
\newblock \bibinfo{title}{Probabilistic {F}orecasts of {W}ind {P}ower
  {G}eneration by {S}tochastic {D}ifferential {E}quation {M}odels}.
\newblock \bibinfo{journal}{Journal of Forecasting} \bibinfo{volume}{35},
  \bibinfo{pages}{189--205}.
\newblock \DOIprefix\doi{10.1002/for.2367}.
\bibitem[{Panik(2017)}]{pani}
\bibinfo{author}{Panik, M.J.}, \bibinfo{year}{2017}.
\newblock \bibinfo{title}{Stochastic {D}ifferential {E}quations: {A}n
  {I}ntroduction with {A}pplications in {P}opulation {D}ynamics {M}odeling}.
\newblock \bibinfo{publisher}{John Wiley \& Sons, Inc.},
  \bibinfo{address}{Hoboken, NJ}.
\newblock \DOIprefix\doi{10.1002/9781119377399}.
\bibitem[{Preston and Wood(2012)}]{prewo}
\bibinfo{author}{Preston, S.}, \bibinfo{author}{Wood, A.T.},
  \bibinfo{year}{2012}.
\newblock \bibinfo{title}{Approximation of transition densities of stochastic
  differential equations by saddlepoint methods applied to small-time
  {I}to--{T}aylor sample-path expansions}.
\newblock \bibinfo{journal}{Statistics and Computing} \bibinfo{volume}{22},
  \bibinfo{pages}{205--217}.
\newblock \DOIprefix\doi{10.1007/s11222-010-9218-8}.
\bibitem[{REN21(2019)}]{ren21}
\bibinfo{author}{REN21}, \bibinfo{year}{2019}.
\newblock \bibinfo{title}{Renewables 2019 {G}lobal {S}tatus {R}eport}.
\newblock \bibinfo{address}{Paris}.
\bibitem[{S{\"a}rkk{\"a} and Solin(2019)}]{saso}
\bibinfo{author}{S{\"a}rkk{\"a}, S.}, \bibinfo{author}{Solin, A.},
  \bibinfo{year}{2019}.
\newblock \bibinfo{title}{Applied {S}tochastic {D}ifferential {E}quations}.
\newblock \bibinfo{publisher}{Cambridge University Press}.
\newblock \DOIprefix\doi{10.1017/9781108186735}.
\bibitem[{Shoji and Ozaki(1998)}]{shoz}
\bibinfo{author}{Shoji, I.}, \bibinfo{author}{Ozaki, T.}, \bibinfo{year}{1998}.
\newblock \bibinfo{title}{Estimation for nonlinear stochastic differential
  equations by a local linearization method}.
\newblock \bibinfo{journal}{Stochastic Analysis and Applications}
  \bibinfo{volume}{16}, \bibinfo{pages}{733--752}.
\newblock \DOIprefix\doi{10.1080/07362999808809559}.
\bibitem[{S{\o}rensen(2012)}]{Sor}
\bibinfo{author}{S{\o}rensen, M.}, \bibinfo{year}{2012}.
\newblock \bibinfo{title}{Estimating functions for diffusion-type processes},
  in: \bibinfo{booktitle}{Statistical {M}ethods for {S}tochastic {D}ifferential
  {E}quations}. \bibinfo{publisher}{Chapman \& Hall/CRC}. volume
  \bibinfo{volume}{124} of \textit{\bibinfo{series}{Monographs on Statistics
  and Applied Probability}}, pp. \bibinfo{pages}{1--107}.
\newblock \DOIprefix\doi{10.1201/b12126}.
\bibitem[{Val{\'e}ry and Gouri{\'e}roux(2011)}]{vago}
\bibinfo{author}{Val{\'e}ry, P.}, \bibinfo{author}{Gouri{\'e}roux, C.},
  \bibinfo{year}{2011}.
\newblock \bibinfo{title}{A quasi-likelihood approach based on eigenfunctions
  for a bounded-valued {J}acobi process (working paper).}
\newblock \bibinfo{howpublished}{Available at
  \url{https://www.researchgate.net/publication/251252253}}.

\end{thebibliography}

\end{document}